\documentclass[twocolumn]{aastex631}

\usepackage{natbib}
\usepackage{amsmath}
\usepackage[normalem]{ulem}

\newcommand{\cdiox}{CO$_2$}
\newcommand{\water}{H$_2$O}
\newcommand{\methane}{CH$_4$}
\newcommand{\oxy}{O$_2$}
\newcommand{\nitro}{N$_2$}
\newcommand{\toaa}{\alpha_\text{TOA}}
\newcommand{\boaa}{\alpha_\text{s}}
\newcommand{\toap}{P_\text{TOA}}
\newcommand{\boap}{P_\text{s}}
\newcommand{\boat}{T_\text{s}}

\shorttitle{Atmospheric Radiative Transfer in Habitable Worlds}
\shortauthors{Simonetti et al.}

\begin{document}

\title{EOS: Atmospheric Radiative Transfer in Habitable Worlds with HELIOS}



\author[0000-0002-7744-5804]{Paolo Simonetti}
\affil{University of Trieste - Dep. of Physics, 
Via G. B. Tiepolo 11, 
34143 Trieste, Italy}
\affil{INAF - Trieste Astronomical Observatory,
Via G. B. Tiepolo 11,
34143 Trieste, Italy}

\author[0000-0001-7604-8332]{Giovanni Vladilo}
\affil{INAF - Trieste Astronomical Observatory,
Via G. B. Tiepolo 11,
34143 Trieste, Italy}

\author[0000-0002-7571-5217]{Laura Silva}
\affil{INAF - Trieste Astronomical Observatory,
Via G. B. Tiepolo 11,
34143 Trieste, Italy}
\affil{IFPU - Institute for Fundamental Physics of the Universe, Via Beirut 2, 34014 Trieste, Italy}

\author[0000-0001-9442-2754]{Michele Maris}
\affil{INAF - Trieste Astronomical Observatory,
Via G. B. Tiepolo 11,
34143 Trieste, Italy}

\author{Stavro L. Ivanovski}
\affil{INAF - Trieste Astronomical Observatory,
Via G. B. Tiepolo 11,
34143 Trieste, Italy}

\author{Lorenzo Biasiotti}
\affil{INAF - Trieste Astronomical Observatory,
Via G. B. Tiepolo 11,
34143 Trieste, Italy}

\author{Matej Malik}
\affil{Dept. of Astronomy, University of Maryland,
College Park, MD 20742, USA}

\author[0000-0002-5312-8070]{Jost von Hardenberg}
\affil{Politecnico di Torino - DIATI,
Corso Duca degli Abruzzi 24,
10129 Torino, Italy}

\begin{abstract}

We present EOS, a procedure for determining the Outgoing Longwave Radiation (OLR) and top-of-atmosphere (TOA) albedo 
for a wide range of conditions expected to be present in the atmospheres of rocky planets with temperate conditions. EOS is based on HELIOS and HELIOS-K, which are novel and publicly available atmospheric radiative transfer (RT) codes optimized for fast calculations with GPU processors. These codes were originally developed for the study of giant planets. In this paper we present an adaptation for applications to terrestrial-type, habitable planets, adding specific physical recipes for the gas opacity and vertical structure of the atmosphere.
To test the reliability of the procedure we assessed the impact of changing line opacity profile, continuum opacity model, atmospheric lapse rate and tropopause position prescriptions on the OLR and the TOA albedo. The results obtained with EOS are in line with those of other RT codes running on traditional CPU processors, while being at least one order of magnitude faster.
The adoption of OLR and TOA albedo data generated with EOS in a zonal and seasonal climate model correctly reproduce the fluxes of the present-day Earth measured by the CERES spacecraft.
The results of this study disclose the possibility to incorporate fast RT calculations in climate models aimed at characterizing the atmospheres of habitable exoplanets.

\end{abstract}

\keywords{astrobiology -- planetary atmospheres -- planetary climates -- radiative transfer simulations -- habitable planets}

\section{Introduction} \label{sec:intro}

Planetary climate simulations are fundamental tools in the search for Earth-like habitable worlds outside the Solar System because, with a proper parameterization of physical quantities, they allow us to estimate the physical conditions in the deeper atmospheric layers and on planetary surfaces from the limited amount of observational data which are typically available for exoplanets.

During the years, many different climate models have been published. In general, they can be categorized either as intermediate/high complexity General Circulation Models (GCMs) or as low complexity Energy Balance Models (EBMs). GCMs perform 3-D hydrodynamical calculations of the atmosphere, taking into account the surface geography and tracking a large number of interactions between different components of the climate system \citep{lin96,hourdin06}. They allow for state-of-the-art predictions but require a large number of input parameters that are usually not known for exoplanets and are very computationally intensive. On the other hand, EBMs make use of simplified relations based on general conservation principles to calculate zonally-averaged quantities \citep[][]{north79,caldeira92,vladilo13}. Despite lacking the accuracy of GCMs and being subject to a set of constraints (for example, on the planetary spin period), they are nonetheless capable of capturing some relevant climate feedbacks, like the ice-albedo feedback. EBMs require a relatively small number of input parameters and far less computational resources, which in turn allow for their use in the rapid exploration of the broad parameters space that characterizes exoplanets \citep[][]{silva17,murante20}.

The planetary energy budget plays the biggest role in determining the general physical state of a planet \citep[see e.g.][]{pierrehumbert11} and it depends on three quantities: the Incoming Shortwave Radiation (ISR, also called Insolation or Instellation), the Outgoing Longwave Radiation (OLR) and the top-of-atmosphere (TOA) albedo, $\toaa$. The ISR depends on the orbital parameters of the planet and the physical characteristics of the host star. The OLR and the TOA albedo, on the other hand, depend on the radiative transfer (RT) properties of the atmosphere (and of the surface, for the latter). ISR and OLR are slightly misleading terms, given that the first one includes also the power emitted by the star as infrared radiation, and the second one may include optical radiation if the considered planet is very hot. 
Anyway, it is evident that the RT plays an essential role in both EBMs and GCMs models.
The modelization of the atmospheric RT is also essential for achieving one of the main goals of present-day astronomy, namely to retrieve the chemical composition of exoplanetary atmospheres from their spectra \citep{mad11,line13}.

There exist a variety of specialized RT codes \citep[][]{briegleb92,clough05} and each of them carries on its task differently, but there are two main strategies: the opacity calculation is performed either on a small number of broad spectral bands via the k-distribution method \citep{goody89}, or line-by-line. In the latter case, the opacity is directly evaluated at each point in frequency while in the former case, a Lebesgue integration on the cumulative distribution function of the opacity inside a band is performed to derive an average value. The advantage of the k-distribution method is computational: in the line-by-line case, the profiles of $10^5-10^6$ lines must be calculated on a very fine frequency grid with potentially $10^7-10^8$ points, which requires a lot of computational power and storage space. However, while being sound and solid, the k-distribution method cannot reach the accuracy of line-by-line calculations. Regardless of the specific strategy used to attain it, the goal of RT codes is to compute the transmission function of the atmosphere, usually dominated by Rayleigh scattering in the optical region and by absorption in the infrared region.

In the past decade, mainly due to the pressure of the consumer electronics market, powerful graphical processing units (GPUs) have become available at a decreasing price, finding applications in a variety of scientific fields where parallelized numerical calculations are used. The use of GPU-based (rather than CPU-based) software can potentially decrease the required computational time by orders of magnitude, and this includes the line-by-line calculations described above \citep[see e.g. the estimates in][]{grimm21}.

For this work, we adapted the GPU-based codes {\tt HELIOS} \citep{malik17,malik19} and {\tt HELIOS-K} \citep{grimm15,grimm21} for studying Earth-like planets. These codes are part of the Exoclimes Simulation Platform (ESP\footnote{\url{https://github.com/exoclime}}) of the University of Bern and were originally devised for the study of hot Jupiter-like planets. {\tt HELIOS-K} is an opacity calculation tool that can operate both in line-by-line and in k-distribution mode, producing opacity functions from line lists provided by repositories such as HITRAN\footnote{\url{https://hitran.org/}} \citep[][]{gordon17} or HITEMP\footnote{\url{https://hitran.org/hitemp/}} \citep[][]{rothman10}. The output of {\tt HELIOS-K} can then be used in {\tt HELIOS}, which is a one-dimensional atmospheric RT code operating in the two-stream approximation. Our goal is to calculate the OLR and the TOA albedo for atmospheric compositions of interest for studies of planetary habitability. In practice, we optimize the calculations for planetary atmospheres with  variable amounts of H$_2$O, CO$_2$, CH$_4$, O$_2$, and N$_2$, taking into account upgraded formulations for the line and continuum opacities. Where these codes lacked the appropriate recipes, we added custom-built scripts. We called this procedure EOS\footnote{Since HELIOS has been named after the Greek god of Sun, for our procedure we chose the name of one of his sisters, the goddess of Dawn.}.

A strategic goal of EOS RT calculations is characterizing the atmospheres of habitable planets for interpreting the exoplanetary spectra that will be collected in the next years.
The immediate goal of EOS is calculating the OLR and TOA albedo in grids of pressure and temperature points to be then incorporated in the Earth-like planets Surface Temperature Model (\texttt{ESTM}), an enhanced EBM with analytical formulations for the meridional transport and a parameterized vertical RT expressed as a function of the surface temperature \citep[][]{vladilo15}. The recalibration of \texttt{ESTM} with the results obtained by EOS, plus a variety of upgraded prescriptions, will be presented in a separate paper \citep[][in preparation]{biasiotti21}. 

The structure of this paper is the following: in the next section we provide a brief overview on atmospheric RT and on the specific challenges offered by CO$_2$ and H$_2$O. In the third section we describe in detail the RT model used in this paper, in particular the prescriptions regarding the vertical structure of the atmosphere employed in our calculations. In the fourth section we test the influence on OLR and TOA albedo of different hypotheses in terms of model variables, opacity prescriptions and vertical thermal structure of the atmosphere. In the fifth section we compare and discuss the OLR and TOA albedo calculated by EOS with the results by other RT models. Finally, in the sixth section, we draw the main conclusions.

\section{Radiative transfer in terrestrial-type planetary atmospheres}
\label{sec:overview}

In this section we will briefly backtrack the path that, from line strengths and shapes obtained from spectroscopic repositories like HITRAN, leads to the solution of the RT equation. We focus on the impact of molecular species such as CO$_2$ and H$_2$O, which are particularly important in habitability studies of terrestrial-type planets.

\subsection{Radiative transfer equations} \label{ss:RT}

The goal of an atmospheric RT model is to solve the RT equation \citep[see][and references therein]{malik17,malik19}:
\begin{equation}
\mu \frac{dI_\lambda}{d\tau_\lambda}=I_\lambda-S_\lambda
\end{equation}
where $\mu$ is the cosine of the incoming radiation incident angle with respect to the normal, $I_\lambda$ is the monochromatic intensity, $\tau_\lambda$ is the optical depth calculated from TOA and $S_\lambda$ is the source function that accounts for both the shortwave radiation scattered into the line of sight and the thermal radiation emitted locally.

The atmosphere is divided into a number of layers and HELIOS calculates the monochromatic fluxes at both interfaces of each atmospheric layer, and in the layer's centre \citep[see Fig.~2 in][]{malik17}. Absorption and non-isotropic scattering are considered both for the incoming and the outgoing radiation. Each layer has an average temperature and an internal linear temperature gradient. The mathematical treatments for the diffuse ($F$) and directional ($F^\text{dir}$) fluxes employed in HELIOS are detailed in \cite{heng14} and \cite{heng18}, respectively. Letting $\mu_\star$ be the cosine of the incoming stellar radiation, the equations for the outgoing (`$\uparrow$') and incoming (`$\downarrow$') fluxes at the lower interface of the $i$th layer are:
\begin{equation}
\label{eq:RT}
F_{i,\uparrow}=\frac{1}{\chi}\biggl(\psi F_{i-1,\uparrow} - \xi F_{i,\downarrow} + 2\pi \epsilon \mathcal{B}_\uparrow + \frac{1}{\mu_\star} \mathcal{I}_\uparrow \biggl)
\end{equation}
\begin{equation*}
F_{i,\downarrow}=\frac{1}{\chi}\biggl(\psi F_{i+1,\downarrow} - \xi F_{i,\uparrow} + 2\pi \epsilon \mathcal{B}_\downarrow + \frac{1}{\mu_\star} \mathcal{I}_\downarrow \biggl)
\end{equation*}
where:
\begin{equation}
\label{eq:RT_supplement}
\mathcal{B}_\uparrow=(\chi+\xi)B_i - \psi B_{i-1} +\frac{\epsilon}{1-\omega_{0}\,g_{0}}(\chi - \xi - \psi) B',
\end{equation}
\begin{equation*}
\mathcal{I}_\uparrow=\psi \mathcal{G}_+ F_{i-1}^\text{dir} - (\xi\mathcal{G}_- + \chi\mathcal{G}_+) F_i^\text{dir},
\end{equation*}
\begin{equation*}
\mathcal{B}_\downarrow=(\chi+\xi)B_i - \psi B_{i+1} +\frac{\epsilon}{1-\omega_{0}\,g_{0}}(\xi - \chi + \psi) B',
\end{equation*}
\begin{equation*}
\mathcal{I}_\downarrow=\psi \mathcal{G}_- F_{i+1}^\text{dir} - (\chi\mathcal{G}_- + \xi\mathcal{G}_+) F_i^\text{dir}
\end{equation*}
and $B$ is the blackbody intensity within the layer. Note that there is a typographical error in Eq.~(9) of \cite{malik19}. In their expression for $\mathcal{B}_\uparrow$, the $-\xi B_{i-1}$ term should be a $-\psi B_{i-1}$. The direct beam flux moves only downward and it is calculated as:
\begin{equation}
F^\text{dir}_i=-\mu_\star F_\star e^{\tau_i/\mu_\star}
\end{equation}
where $F_\star$ is the ISR, which can be modeled as both a blackbody or a real star spectrum, and $\tau_i$ is the optical depth from the TOA to the $i$th layer. $F^\text{dir}_0$ is thus the residual direct beam flux that impacts the planetary surface, of which a fraction $\boaa$ (the surface albedo) is reflected in the form of diffuse radiation and the remainder is absorbed.
The auxiliary quantities used in Eqs.~\ref{eq:RT} and \ref{eq:RT_supplement} are defined as follows:
\begin{equation}
\chi=\zeta_-^2\mathcal{T}-\zeta_+^2
\end{equation}
\begin{equation*}
\xi=\zeta_+\zeta_-(1-\mathcal{T}^2)    
\end{equation*}
\begin{equation*}
\psi=(\zeta_-^2-\zeta_+^2)\mathcal{T}
\end{equation*}
\begin{equation*}
\zeta_\pm=\frac{1}{2}\biggl(1 \pm \sqrt{\frac{1-\omega_{0}}{1-\omega_{0}\,g_{0}}} \biggl)
\end{equation*}
\begin{equation*}
\mathcal{G}_\pm=\frac{1}{2}\biggl[ \mathcal{L} \biggl( \frac{1}{\epsilon} \pm \frac{1}{\mu_\star (1-\omega_{0}\,g_{0})} \biggl) \pm \frac{\omega_{0}\,g_{0} \, \mu_\star}{1-\omega_{0}\,g_{0}} \biggl]
\end{equation*}
\begin{equation*}
\mathcal{L}=\frac{(1-\omega_{0}) (1-\omega_{0}\,g_{0}) - 1}{1/\mu_\star^2 - 1/\epsilon^2 (1-\omega_{0}) (1-\omega_{0}\,g_{0})}
\end{equation*}
\begin{equation*}
B'=\frac{\Delta B}{\Delta \tau}
\end{equation*}
In order to calculate these quantities, four physical variables must be evaluated in each layer, namely $\epsilon$, $\omega_0$, $g_0$ and $\Delta \tau$. The first one, $\epsilon$, is called first Eddington coefficient and is defined as the ratio between the first and second moment of the intensity. It describes how much diffuse is the radiation, spanning the range from 0.5 (fully isotropic) to 1.0 (fully directional on the vertical axis). The second variable, $\omega_0$, is the single-scattering albedo and is defined as $k_\text{sca} / (k_\text{sca}+k_\text{abs})$, where $k_\text{sca}$ and $k_\text{abs}$ are the molecular cross sections of the gas mixture for Rayleigh scattering and absorption, respectively. The third variable, $g_0$, is the asymmetry parameter and describes the preferred direction in which radiation is scattered. It spans the range from -1 (fully backward scattering) to +1 (fully forward scattering); if $g_0=0$ the scattering is isotropic. Finally, $\Delta \tau$ is the optical thickness of the layer, which is also employed to calculate the transmission function: $\mathcal{T}$:
\begin{equation}
\mathcal{T}=\exp\big[{-\epsilon^{-1} \sqrt{(1-\omega_0\,g_0)(1-\omega_0)} \Delta \tau}\big]
\end{equation}
The optical thickness is the product of
the scattering and absorption cross sections of the gas mixture multiplied by the columnar density $u$:

\begin{equation}
\tau = (k_\text{sca}+k_\text{abs}) \, u = (k_\text{sca}+k_\text{abs}) \int_{0}^{l} \rho(l') ~dl'
\end{equation}
where $\rho$ is the volumetric density along the geometrical optical path $l$ inside the layer.

In EOS, $\omega_0$ and $\Delta \tau$ are calculated on the spot while $\epsilon$ and $g_0$ are kept constant. In particular, we chose $\epsilon=0.5$ and $g_0=0$. The first assumption is justified by the fact that we are dealing with diffuse radiation, both in terms of thermal emission from the planet and the atmosphere (for the calculation of the OLR) and in terms of the light scattered by the air or reflected by the surface (for the calculation of the TOA albedo). This is consistent with the hemispheric mean closure \citep{toon89}. The second assumption is justified by the fact that, in our tests, we considered only the Rayleigh scattering by gas molecules and not the Mie scattering by particulate. In other words, we computed clear-sky results, without considering the effects of clouds or suspended dust.
The cloud contributions to the albedo and OLR are accounted for in the climate model. A brief description of this procedure for the case of the ESTM can be found in Sec.~\ref{ss:implications}; full details are described in \cite{vladilo15} and in \cite{biasiotti21}.


\subsection{Scattering cross sections}

Since we run clear-sky RT calculations, the scattering opacity of the atmosphere is considered to be entirely caused by the Rayleigh scattering on gas molecules. As described in \cite{sneep05}, the cross section for this process is:
\begin{equation}
k_\text{sca}(\lambda)=\frac{24\pi^3}{n_\text{ref}^2\,\lambda^4}\biggl(\frac{\Tilde{n}_\lambda^2-1}{\Tilde{n}_\lambda^2 + 2}\biggl)^2\,K_\lambda
\end{equation}
where $\Tilde{n}_\lambda$ is the wavelength-dependent refractive index, $n_\text{ref}$ is the number density at which the refractive index has been calculated and $K_\lambda$ is the King depolarization correction factor \citep{king23}.

The functional forms for $\Tilde{n}_\lambda$ and $K_\lambda$ are taken from the literature. For \nitro, \oxy~and \cdiox~we referred to \cite{sneep05} and \cite{thalman14}, and for \water~to \cite{murphy77}, \cite{schiebener90} and \cite{wagner08}. We did not consider the Rayleigh scattering on \methane~since it has very low concentrations (of the order of one part per billion in volume) or it is absent altogether in the atmospheres that we tested.

\subsection{Absorption cross sections}

The absorption cross section (also called absorption coefficient) for an individual line centered at frequency $\nu_0$ can be calculated at each frequency $\nu$ as follows:
\begin{equation}
k_\text{abs}(\nu)=Sf(\nu-\nu_0,\gamma)
\end{equation}
where $S$ is the line strength, $f(\nu_0-\nu,\gamma)$ is the line shape function and $\gamma$, using a formalism employed also in the HITRAN database, takes into account the effects of line broadening due to pressure and temperature effects. There exist many different models for the line shape function \citep[e.g.][]{galatry61,rautian67,ngo13}, but for atmospheric RT models a Voigt profile is usually adopted \citep[e.g. in \texttt{LBLRTM}, see][]{clough05}. The Voigt function is obtained by the convolution of Gaussian and Lorentzian distributions, and reduces to the Lorentzian distribution when the distance from the line centre is larger than some half-widths. Therefore, in such cases, the absorption coefficient (neglecting the line centre pressure-induced shift) can be written as:
\begin{equation}
\label{eq:kappa}
\begin{split}
k_\text{abs}(\nu)=Sf_L(\nu-\nu_0,\gamma)=\\
=\frac{S}{\pi}\frac{\gamma(P,T)}{\gamma(P,T)^2+(\nu-\nu_0)^2}
\end{split}
\end{equation}
The factor $\gamma$, which is the half-width at half-maximum (HWHM) of the Lorentz distribution, is then calculated as:
\begin{equation}
\label{eq:gamma}
\begin{split}
\gamma(P,T) &= \gamma_\text{nat}+\gamma_\text{self}(P,T)+\gamma_\text{for}(P,T)= \\
&=\frac{A}{4\pi c}+ \biggl(\frac{T}{T_{\text{ref}}}\biggl)^{-n}\biggl(\tilde{\gamma}_\text{self}(P_\text{ref},T_\text{ref})\frac{P_{\text{self}}}{P_{\text{ref}}}+ \\
&+\tilde{\gamma}_\text{for}(P_\text{ref},T_\text{ref})\frac{P-P_{\text{self}}}{P_{\text{ref}}}\biggl)
\end{split}
\end{equation}
The first term represents the natural width of the line (being $A$ the Einstein coefficient and $c$ the speed of light), while the second and the third terms represent the pressure broadening due to collisions with molecules of the same species and collisions with molecules of other species, respectively. These last two terms are given for a reference temperature $T_\text{ref}$ and pressure $P_\text{ref}$ and scaled according to the equation. in addition, the exponent $n$ is empirical.

It must be noted that, in theory, each couple of molecules has its own $\tilde{\gamma}$ value, and as such the $\tilde{\gamma}_\text{for}$ term should really be a summation over all the possible binary combinations of species in the considered mixture, weighted by their molar fractions. However, even the calculation for a relatively simple mixture of gas would require a large number of laboratory measurements that are not currently available. Therefore, publicly available repositories usually give only a single value for $\tilde{\gamma}_\text{for}$, derived in the case of dry Earth air. This is the case, for example, of the standard HITRAN tables \citep{rothman05} read by {\tt HELIOS-K}. 

In this work, the solution of the RT equation via the two stream approximation and the calculation of the optical depth of the gas mixture for a given composition is performed by {\tt HELIOS}. On the other hand, the calculation of the absorption coefficients for each chemical species is handled by {\tt HELIOS-K}.

In the rest of this section we discuss the main physical recipes adopted in EOS in order to adapt {\tt HELIOS} and {\tt HELIOS-K} to the study of the atmospheres of habitable planets. Following classic studies of circumstellar habitability \citep[][]{kasting84} we focus on Earth-like atmospheres with variable amounts of water vapor, and \cdiox-dominated atmospheres. The former are related to the emergence of the runaway greenhouse feedback and are important for the study of the inner edge of the circumstellar habitable zone (CHZ), whereas the latter are relevant for the study of the outer edge of the CHZ.
In this context it is essential to investigate the opacity features of water and carbon dioxide.

\subsection{Carbon dioxide line wings} \label{ss:CO2wings_theory}

One problem with the use of Voigt profiles is the well-known sub-Lorentzian behaviour of the far line wings of \cdiox ~\citep[][]{winters64}: above a certain distance from the line centre, the opacity of the \cdiox~absorption lines is lower than what expected by applying Eq.~\ref{eq:kappa}. A common way to model the deviation from the ideal line shape consists in multiplying the Lorentz distribution $f_L$ by some other exponentially-decreasing $\chi(\lvert \nu-\nu_0 \rvert,P,T)$ function for distances from the line centre above a given threshold \citep[$\nu_c$,][]{burch69}. Letting $\lvert \nu-\nu_0 \rvert = \Delta\nu$, for a single-component gas this takes the following form:
\begin{equation}
\label{eq:chi}
f(\Delta\nu,P,T)=
\begin{cases}
f_L(\Delta\nu,P,T) & \Delta\nu < \Delta\nu_c \\
f_L(\Delta\nu,P,T)\chi(\Delta\nu,T) & \Delta\nu \ge \Delta\nu_c \\
\end{cases}
\end{equation}
Whereas for a mixture of gases the $\nu \ge \nu_c$ case becomes:
\begin{multline}
\label{eq:PH89}
f(\Delta\nu,P,T)=\\
=\frac{1}{\pi}\frac{\gamma_\text{self}(P,T)\chi_\text{self}(\Delta\nu,T)+\gamma_\text{for}(P,T)\chi_\text{for}(\Delta\nu,T)}{\gamma(P,T)^2+\Delta\nu^2}
\end{multline}
where $\gamma$ is defined as in Eq.~\ref{eq:gamma} and $\gamma_\text{nat}$, which for Earth-like conditions is much smaller than both the pressure induced terms, has been dropped in the numerator. The critical distance from the line centre $\Delta\nu_c$ have usually a value of 3 cm$^{-1}$ \citep{burch69,perrin89}.

Many different $\chi(\Delta\nu,T)$ functions have been proposed, both empirical and theoretical \citep{ma99}. A widely known empirical formulation is the one proposed by \citet{perrin89}, hereafter PH89, who experimentally studied the sub-Lorentzian behaviour of the \cdiox ~lines in the $4.3 ~\mu m$ band for a large interval of temperatures and pressures. They proposed the following 4-case function: 
\begin{multline}
\chi(\Delta\nu,T)=\\
\begin{cases}
1 & \Delta\nu \le \sigma_1 \\
\exp(-b_1(\Delta\nu-\sigma_1)) & \sigma_1 < \Delta\nu \le \sigma_2 \\
\exp(-b_1(\sigma_2-\sigma_1)-b_2(\Delta\nu-\sigma_2)) & \sigma_2 < \Delta\nu \le \sigma_3 \\
\exp(-b_1(\sigma_2-\sigma_1)-b_2(\sigma_3-\sigma_2)+\\
-b_3(\Delta\nu-\sigma_3)) & \Delta\nu > \sigma_3 \\
\end{cases}
\end{multline}
where $\sigma_1=\Delta\nu_c$ and the $b_i$ terms contain the temperature dependence as:
\begin{equation}
b_i(T)=\alpha_i+\beta_i\exp(-\epsilon_i T)
\end{equation}
The authors provide the values for $\alpha_i$, $\beta_i$, $\epsilon_i$ and $\sigma_i$ for \cdiox-\cdiox ~and \cdiox-N$_2$ mixtures.

Other groups \citep{bezard90,pollack93,tonkov96} used similar schemes but found different values. Such discrepancies are caused by the fact that lines in different spectral bands are more or less sub-Lorentzian than in the $4.3 ~\mu m$ band. A source of uncertainty frequently cited by experimentalists is the possible contamination by other optically active gases with strong absorption features in the studied spectral regions. Finally, there is the fact that not all the research groups in the field have derived these factors for different points in the pressure-temperature space. For example, \cite{tonkov96}, hereafter T96, conducted their observations at room temperature only (296 K).

Another aspect that must be considered is the wing truncation distance from the line centre. This is less of an issue when a sub-Lorentzian $\chi$ function is applied with respect of the pure Voigt case, but this choice can still influence the final output \citep{haus10}. Proposed values range from 120 cm$^{-1}$ to no truncation whatsoever, but \cite{wordsworth10} verified that choosing 500 cm$^{-1}$ even for pure Voigt profiles produced only a negligible difference with respect to untruncated spectra, while being computationally more efficient.

For \cdiox-dominated atmospheres we adopt the prescriptions of PH89, while for Earth-like atmospheres we keep a Voigtian profile truncated at 25 cm$^{-1}$, as reported in Table  \ref{tab:models}. In Sec.~\ref{sec:input} we will show how different sub-Lorentzian far wings and truncation prescriptions impact the final results for a \cdiox-dominated atmosphere. This will help to check these kinds of uncertainties inherent to RT models and how they reflect on climate simulations.

\subsection{Continuum opacity} \label{ss:continuum}

Besides the molecular transitions described by groups of spectral lines, different chemical species also have absorption continua that vary slowly in frequency.
The contribution to the total opacity that comes from these continua is particularly important in the spectral windows, i.e.~in the regions of the electromagnetic spectrum where there are few or no allowed transitions.
They are characterized by a quadratic dependence on pressure, which makes them the principal contributor to a given species' opacity in mid- and high-pressure environments.
The mechanisms responsible for the continuum absorption are not fully understood, and at least three different theories have been proposed and applied to different specific cases.

The first one explains the continuum opacity as the product of collision-induced dipole absorption, usually abbreviated in CIA. This is commonly invoked for non-polar molecules such as \cdiox ~\citep{gruszka97,gruszka98} or N$_2$ \citep{borysow86}, while it has been criticized when applied to polar species \citep[see e.g.][]{vigasin14}.

The second theory considers the continuum as produced by stable and meta-stable dimers formed during molecular collisions. It is usually applied to polar molecules such as water \citep{ptashnik11}, but has also been used to explain the absorption in the 1100-1800 cm$^{-1}$ region of \cdiox ~\citep{baranov99,baranov04}. Strong support for this explanation, at least in the \water ~case, comes from the partial resolution of dimers' spectrum in the microwave region \citep{tretyakov13,serov14}.

The third possibility explains the continuum as the sum of super-Lorentzian mid-wing (10-100 cm$^{-1}$ from the line centre) and sub-Lorentzian far-wing contributions from nearby allowed transitions. This has also formed the basis for the original Clough-Kneizys-Davies (CKD) continuum model of \water ~\citep{clough89}, even though it is more of a practical approach, rather than a physical explanation.

In regard to the first two explanations, \cite{shine16} also note that it can be difficult to clearly separate the contributions of CIA from those of dimers, especially in the absence of a full understanding of the quantum chemistry of bound and quasi-bound molecular complexes, at least for water.

In general, continuum opacity can be modeled as the second-order contribution in density to the monochromatic optical depth. As such the total absorption coefficient will then be
\begin{multline}
k_\text{abs}(\nu) = k_\text{lin}(\nu) + k_\text{cnt}(\nu,n) =\\
=k_\text{lin}(\nu) + \tilde{k}_\text{cnt}^\text{self}(\nu,T)n_\text{self} + \tilde{k}_\text{cnt}^\text{for}(\nu,T)n_\text{for}
\end{multline}
where $k_\text{lin}(\nu)$ is the absorption due to the allowed transitions modeled as in Eq.~\ref{eq:kappa} and $k_\text{cnt}(\nu,n)$ is the molar density-dependent continuum contribution. As in the case of pressure broadening and the $\chi$ function, there is a distinction between self-induced continuum and the (usually smaller) foreign-induced continuum. Once again, a precise absorption model should consider all the possible binary combinations of molecules in the gas mixture, but this is usually not possible given the lack of data. HITRAN offers an increasing number of tabulated values of self- and foreign-continuum opacities for a selection of species. The specific recipes adopted in this work are reported in the following subsections.

\subsubsection{Carbon dioxide continuum} \label{ss:CO2cont}

Foundational works on early-Mars and early-Earth climates such as \cite{kasting84} and \cite{kasting91}, hereafter K91, used a parameterized continuum model derived by \cite{pollack80} and tested against microwave and far infrared observations made by the Pioneer Venus spacecraft. This model has been deemed to be too opaque in the 250-500 cm$^{-1}$ region and too transparent in the 1300-1400 cm$^{-1}$ region \citep{wordsworth10}.

More recent data on the \cdiox ~self-continuum, as measured by laboratory experiments, numerical simulations and Venus observations, have been collected by HITRAN. The self-continuum is tabulated for four different bands: 0-750 cm$^{-1}$ \citep{gruszka97,gruszka98}, 1000-1800 cm$^{-1}$ \citep{baranov99}, 2510-2850 cm$^{-1}$ \citep{baranov03} and 2850-3250 cm$^{-1}$ \citep{baranov18}. Data for each band do not span the same interval in temperature. For example, the continuum absorption for the first of these bands is given for 10 points in the 200-800 K range, while for the last one only a single point at 298 K is available. In this paper, given the non-negligible dependence of the continuum opacity on temperature, and for a more meaningful comparison with the results obtained by \cite{wordsworth10}, we have chosen to use only the 0-750 cm$^{-1}$ and the 1000-1800 cm$^{-1}$ bands continua. As in that work, we will label this model of continuum as ``GBB'' (Gruszka-Borysow-Baranov).

\subsubsection{Water vapor continuum} \label{ss:H2Ocont}

As shown, for example, in \citet[][hereafter KOP13]{kopparapu13a,kopparapu13b}, correctly accounting for the water vapor continuum is of paramount importance for a meaningful modeling of Earth-like planets and for climate predictions on our own planet. An early model adopted in \cite{kasting88}, used also in \cite{kasting93}, is based on the Fels-Goody random band model \citep{fels79,goody64}, computed on synthetic spectra published in \cite{afgl71}.

\cite{clough89} developed a super-Lorentzian treatment to account for the continuum opacity of the \water ~absorption lines, called the CKD model. More recently, \cite{mlawer12} expanded this model to incorporate the effects of the foreign-induced continuum, in what is known as the MT-CKD model. Both the CKD and the improved MT-CKD models are widely used in a variety of climate and weather simulations. Another formulation, prominently used in KOP13, is the BPS model by \cite{paynter11}. This diverges slightly from the MT-CKD model on a conceptual basis because the continuum absorption comes directly from empirical data and is independent of the analytical formulation of line profiles.

For this work we have adopted the latest version (3.4) of the MT-CKD model, where the continuum of water vapor is calculated by the \texttt{CNTNM} code, which is a standalone version of the continuum calculation module of the \texttt{LBLRTM}. The continuum is tabulated in the 0-10000 cm$^{-1}$ frequency range and in the 200-400 K temperature range.


\section{Calculation of the atmospheric radiative fluxes}
\label{sec:model}

In the EOS scheme, the long- and short-wavelength fluxes at the top of the planetary atmosphere are calculated under the reverse procedure described by \cite{kasting93}.
In practice, instead of calculating the surface temperature as a function of insolation, TOA fluxes are evaluated for a given surface temperature, lapse rate and atmospheric composition. 
By iterating this procedure for a grid of values of surface temperatures and atmospheric parameters, we generate radiative transfer tables that can be interfaced to climate models, which find the insolation that is capable to balance these energy fluxes.
This reverse procedure is commonly adopted in climate calculations of the habitable zone (as e.g.~in KOP13) and, in particular, by \texttt{ESTM} \citep{vladilo15}. For these applications two relations need to be calculated: (i) the OLR as a function of $T_s$ and (ii) the TOA albedo as a function of $T_s$, zenith angle, $z$, and surface albedo, $\boaa$. These quantities are evaluated for a fixed\footnote{Specifically, the columnar masses of the non-condensable species are kept constant. On the other hand \water, which is condensable, varies as a function of temperature (see Sec.~\ref{ss:earth_lapserate}).} atmospheric composition and a chosen P-T vertical structure. For each atmospheric structure and composition we construct two tables, one for the OLR($T_s$) and another for the $\toaa$($T_s$,$z$,$\boaa$), and this procedure is repeated for hundreds of \texttt{HELIOS} runs in order to cover the parameter space of interest for terrestrial-type planets. In a second moment, \texttt{ESTM} uses the content of these tables to actually calculate the energy balance of the planet.
The technical details and the step-by-step procedure about the construction of the tables are reported in the Appendix.

\begin{figure*}[]
\plottwo{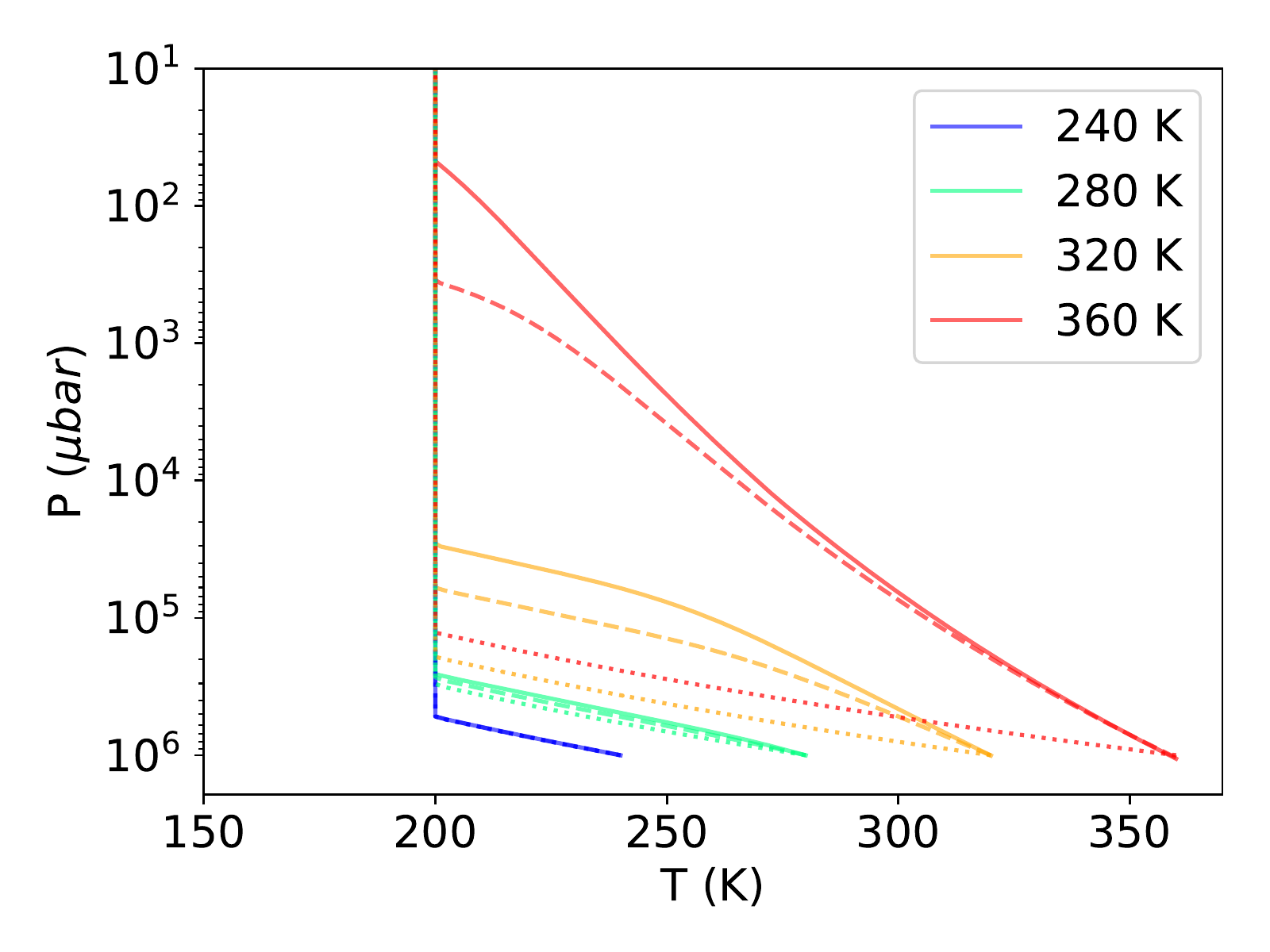}{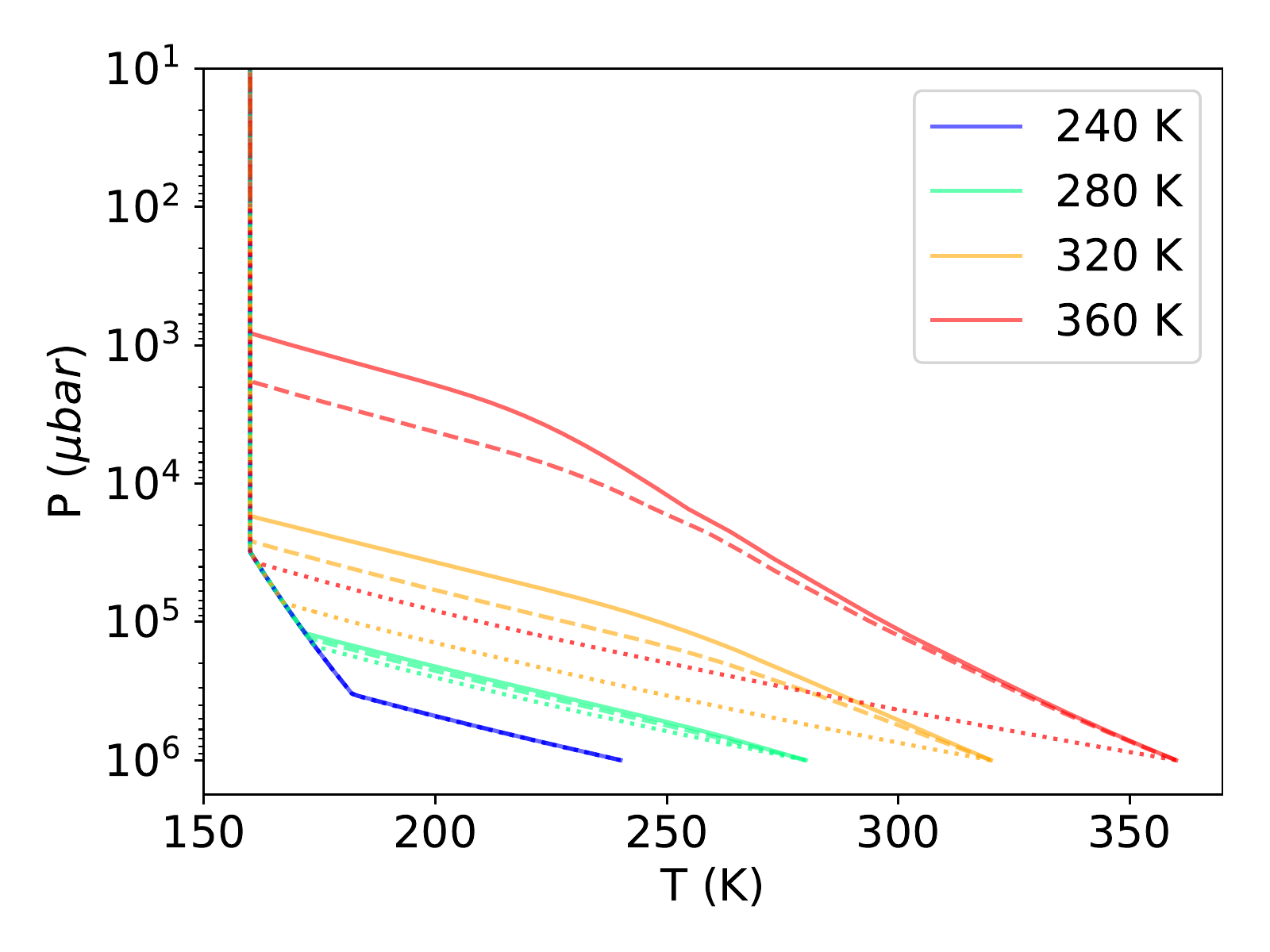}
\caption{Vertical P-T profiles adopted for the two sets of atmospheres considered in this work. Left: profiles for an Earth-like atmosphere with a 200 K tropopause. Right: profiles for a \cdiox-dominated atmosphere with a 160 K tropopause. Dotted, dashed and solid lines refer to dry, 60\% moist and 100\% moist (in relative humidity) adiabatic lapse rates, respectively. Different colors refer to different surface temperatures as indicated in the legend. To allow for an easier comparison between lapse rates, the $\boap$ is set equal to 1 bar regardless of the relative humidity, while in the tests we actually accounted for the increase of the surface pressure due to enhanced evaporation. 
\label{fig:lapserates}}
\end{figure*}

The vertical structure of the atmosphere has a strong impact on the OLR and TOA albedo of a planet and must be considered carefully. In general, we have considered variable lapse rate tropospheres topped by isothermal stratospheres. The bottom-of-atmosphere (BOA) pressure $\boap$ has been varied both directly (i.e.~in its non-condensing component) and indirectly (i.e.~to account for the increasing partial pressure of \water ~when the surface warms up). In all models the TOA pressure $\toap$ (the pressure of the uppermost simulated layer) has been set equal to 1 $\mu$bar. For each atmosphere considered, we adopted a constant chemical composition along the pressure axis, except when condensation occurs. 
Following the standard \texttt{HELIOS} procedure we have run the code using a 10 layers per order of magnitude log-spaced pressure grid, which corresponds to a total of 60 layers in most cases.

\subsection{Earth-like atmospheres}
\label{ss:earth_lapserate}

For Earth-like tropospheres, we adopted the classical two-component moist pseudoadiabat\footnote{Strictly speaking the process is not reversible (and hence not adiabatic), because when the minor component condenses, it is removed from the atmosphere via precipitation.}
(referred to as \textit{adiabat} in the rest of this paper), where the main component is noncondensable (identified by the subscript "n") and the minor component is condensable (identified by the subscript "v" for vapor). Both of them are thought to behave ideally, which is acceptable for Earth-like pressure and temperature conditions. 
We used Eq.~(2.33) of \cite{pierrehumbert11}:
\begin{equation}
\frac{d\log T}{d\log P}=\frac{R}{m_n C_{p,n}}\frac{1+\frac{L m_n r_v(T)}{RT}}{1 + \big(\frac{C_{p,v}}{C_{p,n}}+\frac{L}{C_{p,n}T}\big(\frac{L m_v}{R T}-1\big)\big)r_v(T)}
\label{eq:Earthlapserate}
\end{equation}
where the $C_p$ are the specific isobaric heats, the $m$ are the molar masses, $R$ is the ideal gas constant, $L$ is the latent heat of the condensable and $r_v$ is the mass density ratio between condensable and noncondensable components. This last term depends on temperature, since the density of the condensable is a function of temperature via the Clausius-Clapeyron relation.
The results for different surface temperatures $\boat$ can be seen in Fig.~\ref{fig:lapserates}, left-hand panel, as solid lines. 

Alongside the standard moist lapse rate derived from thermodynamics, we have also considered a slightly modified version where the \water ~partial pressure in the $r_v$ term has been scaled with a relative humidity lower than 100\%. In theory, this should not be possible: if air is not saturated, water cannot condense and the atmosphere should have a dry adiabatic lapse rate. However, the real average lapse rate on Earth is neither moist nor dry, but somewhere in-between. This is caused by a large number of factors, the most relevant of which is the general inhomogeneity of the real terrestrial atmosphere \citep{pierrehumbert07}. Given that (i) our intent here is to use the results from \texttt{HELIOS} in an EBM model that cannot capture such small scale variations, (ii) the non-radiative vertical energy flux in an Earth-like atmosphere is non-negligible and (iii) \water ~condensation plays an important role in this flux, we modified the moist adiabat profile by re-scaling the partial pressure of water vapor according to the relative humidity. As one can see in Fig.~\ref{fig:lapserates} (dashed lines in  left-hand panel), this change has the effect of steepening the pressure-temperature (P-T) relation in the troposphere, giving it a shape that is in-between those of the theoretical moist and dry adiabatic models (dotted lines in the same panel). The strongest effect of this change is felt at high $\boat$ ($> 320$ K). 

As far as the isothermal stratosphere temperature is concerned, following the literature we adopted a value equal to 200 K, representative of the modern Earth.
In Sec.~\ref{ss:atmopT} we discuss the impact on the results of different choices of the P-T profiles and of the stratospheric temperature.

\subsection{CO$_2$-dominated atmospheres}
\label{ss:co2_lapserate}

For \cdiox-dominated tropospheres, we have adopted the model detailed in appendix A of K91. In this case, the troposphere is divided into two sections: (i) a lower troposphere where only the minor component (i.e.~\water) condenses and (ii) an upper troposphere where both the minor and the major (i.e.~\cdiox) condense. The lapse rate for the lower troposphere, as per Eq.~(A18) of K91, can be written as:
\begin{equation}
\frac{dP}{dT}=\frac{C_{p,n} + r_v \big(C_c + \frac{dL}{dT} + \frac{L}{T} \big(\frac{m_vL}{RT} + \frac{T}{V} \big(\frac{\partial V}{\partial T}\big)_P\big)\big)}{\frac{T}{m_n}\big(\frac{\partial V}{\partial T}\big)_P - \frac{r_vL}{V} \big(\frac{\partial V}{\partial P}\big)_T}
\label{eq:CO2lapserate}
\end{equation}
where $C_c$ is the specific heat of the condensed phase of the condensable and $V$ is the molar volume. As for the derivation of Eq.~\ref{eq:Earthlapserate}, the condensable is supposed to immediately precipitate out of the atmosphere when formed. The subscripts "P" and "T" indicate that the relevant partial derivatives have been taken at constant pressure and temperature, respectively. This formulation allows us to include the effects of a non-ideal behaviour of the major atmospheric constituent on the tropospheric P-T structure. In our case, this is necessary to properly treat relatively dense and/or cold atmospheres, as can be expected near the outer edge of the CHZ. Thermodynamic quantities relevant to this calculation have been taken from the reference tables produced by the National Institute of Standards and Technology (NIST) and based on the work of \cite{span96}. 

The lapse rate of the upper troposphere follows the condensation curve of the \cdiox, which is the major constituent. It can be calculated in this way because of the very low partial pressure of water vapor at the typical temperatures of the upper troposphere, and therefore $\text{P}_\text{tot}=\text{P}_\text{major}+\text{P}_\text{minor} \simeq \text{P}_\text{major}$. Once again, we have used tabulated data from the NIST. We studied also a case with a dry \cdiox ~atmosphere, in which we simply set $r_v$ to zero. In that case, Eq.~\ref{eq:CO2lapserate} reduces to Eq.~(A19) in the Appendix of K91.

To derive the stratospheric temperature of \cdiox-dominated atmospheres,  K91 run a radiative-convective equilibrium calculation of a specific case (that with a $\boap=0.35$ bar), choosing the cold-trap temperature and then scaling that result for the other tested cases using the skin temperature\footnote{The temperature of the optically thin part of the atmosphere.} relation, which is a function of the ISR and TOA albedo.
Given the increased scattering in thicker atmospheres, the skin temperature decreases when surface pressure increases. However, in that work it is shown that the skin temperature variations (i) are small ($\sim 10$ K increasing $\boap$ from 0.35 to 5 bars) and (ii) have little influence on the surface temperature. Therefore, for \cdiox-dominated atmospheres we decided to fix the stratospheric temperature to 160 K, a choice made also by \cite{halevy09}.

\section{Results}
\label{sec:input}

In this section we show how the OLR and TOA albedo calculated with our procedure are affected by variations of simulation parameters and physical input variables. We begin by studying the influence of two computational parameters, namely the number of layers and the TOA pressure. Then we present the effects of changing the lapse rate formulation and the tropopause prescription. Finally we show the effects of changing chemical composition. For \cdiox, we also explore the influence of different sub-Lorentzian line shape prescriptions.
In Table \ref{tab:models} we list the name and main features of the test cases discussed below.
For a given atmosphere, the OLR is studied as a function of $T_s$ and the TOA albedo is studied as a function of $T_s$ and $z$. The dependence of TOA albedo on $T_s$ is relevant in moist atmospheres owing to the presence of \water~absorption bands in the near infrared. The increased amount of \water~at higher $T_s$ thus reduces the amount of light scattered, in turn reducing the TOA albedo.
We remark that expressing both the OLR and the TOA albedo as a function of $T_s$ is required by the reverse procedure, as described at the beginning of Sec.~\ref{sec:model}.

\begin{table*}[]
\centering
\begin{tabular}{lccccccc}
\hline
Name & \nitro & \oxy & \cdiox & \methane & \water & \cdiox~line & \cdiox~CIA\\
\quad & fraction & fraction & fraction & fraction & RH & presc. & presc.\\
\hline
Pre-industrial Earth  & 0.79  & 0.21 & 280$\times$10$^{-6}$  & 0.6$\times$10$^{-6}$  & 60\% & Voigt, 25 cm$^{-1}$ & no \\
Modern Earth 1 & 0.79  & 0.21  & 360$\times$10$^{-6}$  & 1.8$\times$10$^{-6}$  & 50\% & Voigt, 25 cm$^{-1}$ & no \\
Modern Earth 2 & 0.79  & 0.21  & 360$\times$10$^{-6}$  & 1.8$\times$10$^{-6}$  & 60\% & Voigt, 25 cm$^{-1}$ & no \\
Saturated Modern Earth  & 0.79 & 0.21 & 360$\times$10$^{-6}$  & 1.8$\times$10$^{-6}$  & 100\% & Voigt, 25 cm$^{-1}$ & no\\
Post-industrial Earth  & 0.79  & 0.21  & 1152$\times$10$^{-6}$  & no & 60\% & Voigt, 25 cm$^{-1}$ & no\\
Dry CO$_2$-dominated  & no & no & 1.0 & no & no & Voigt, 25 cm$^{-1}$ & GBB\\
Dry CO$_2$-dominated  & no & no & 1.0 & no & no & PH89, 500 cm$^{-1}$ & GBB\\
Moist CO$_2$-dominated  & no & no & 1.0 & no & 100\% & Voigt, 25 cm$^{-1}$ & GBB\\
Moist CO$_2$-dominated  & no & no & 1.0 & no & 100\% & PH89, 500 cm$^{-1}$ & GBB\\
Moist CO$_2$-dominated  & no & no & 1.0 & no & 100\% & T96, 350 cm$^{-1}$ & GBB\\
\hline
\end{tabular}
\caption{A summary of the test cases studied in this paper. The \cdiox-dominated models have been tested for different CO$_2$ partial pressures. The total surface pressure has been varied according to the temperature dependence of the \water~partial pressure. Entries in the the fourth and fifth columns are in volume fractions. The sixth column gives the relative humidity. The last column reports the prescriptions adopted for the \cdiox~line opacity.}
\label{tab:models}
\end{table*}

\subsection{Changes in computational variables} \label{ss:atmogrid}

\begin{figure*}[]
\plottwo{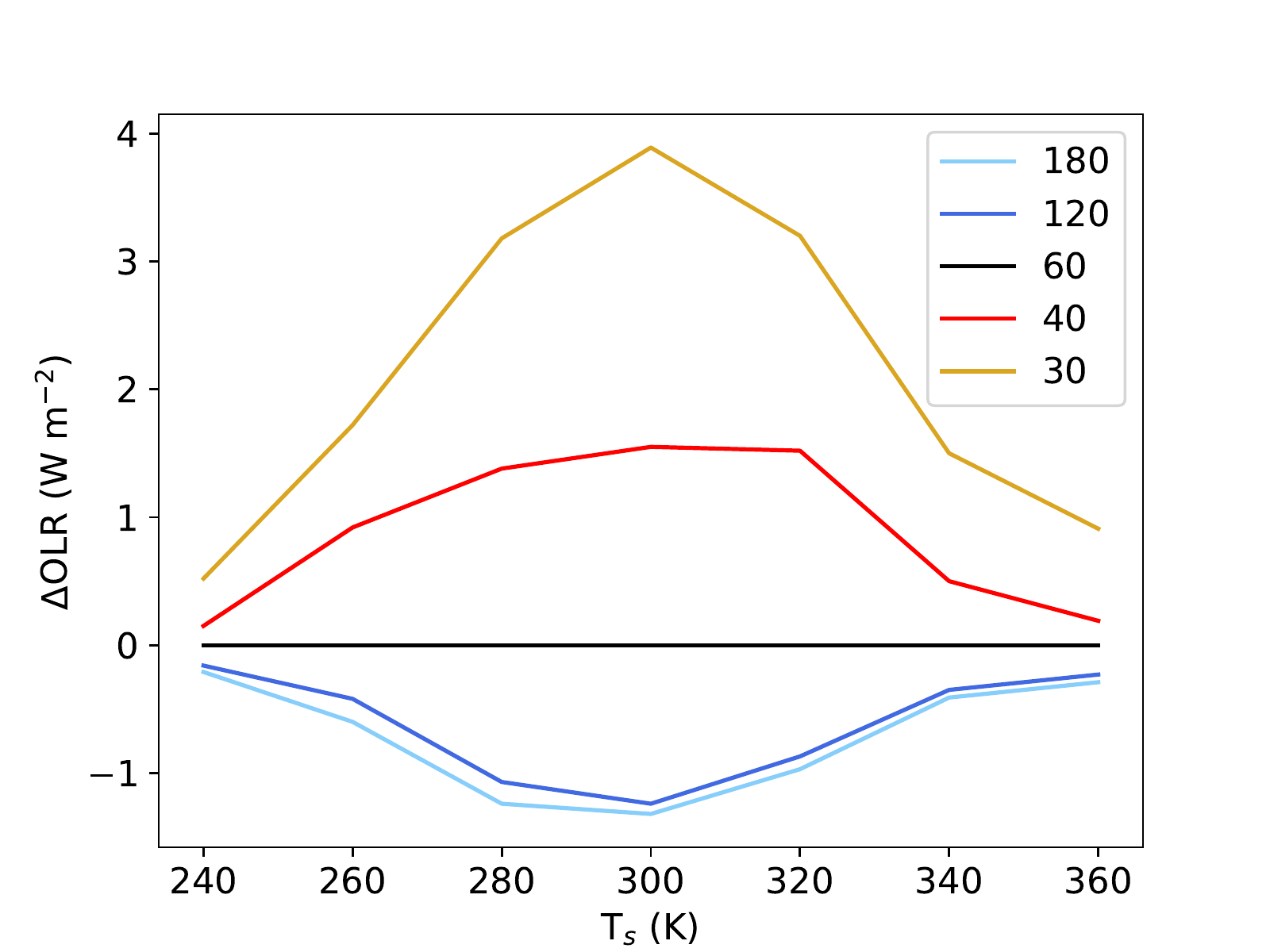}{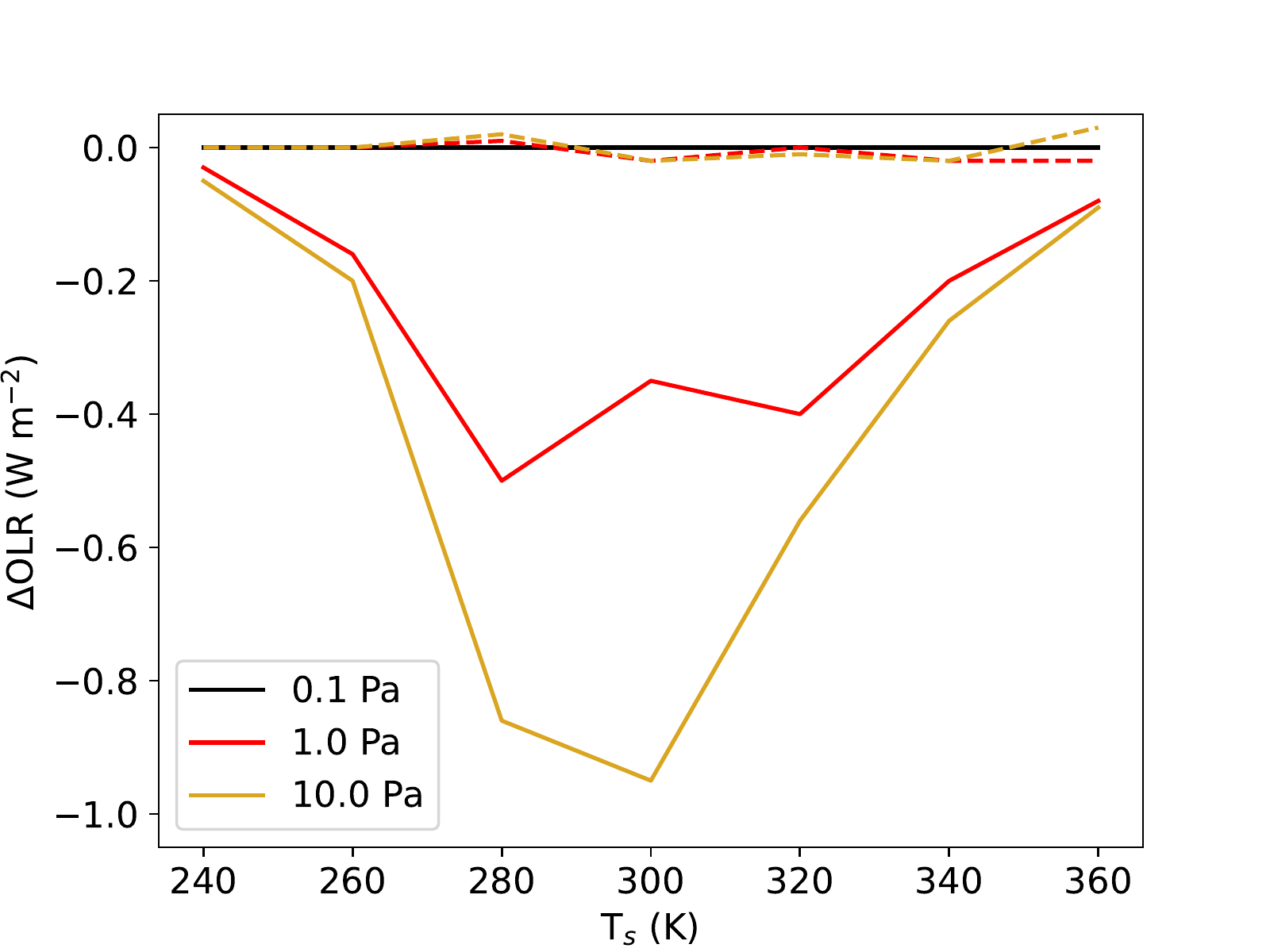}
\caption{The absolute deviation (in W m$^{-2}$) from the OLR of our reference model for Earth atmosphere (in black) as a function of temperature. On the left hand panel we show models with different layer densities, reported in the legend as the total number of layers used each simulation. 
On the right hand panel models with different TOA pressures are shown. Solid lines refer to models with a fixed number of total layers (60 in this case), while dashed lines refer to models with a fixed layer density (10 per order of magnitude in this case). \label{fig:atmogrid}}
\end{figure*}

\begin{figure*}[]
\plottwo{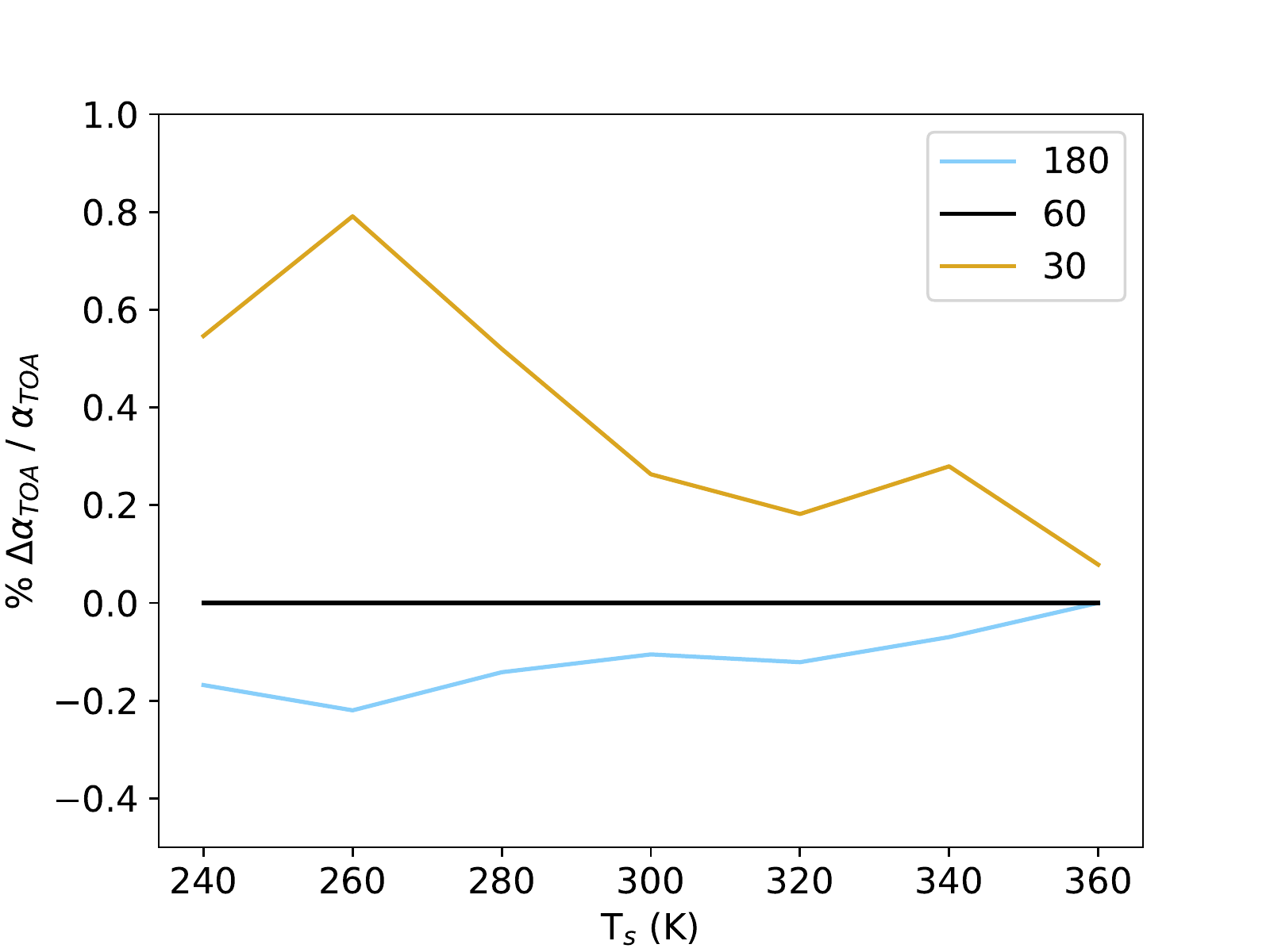}{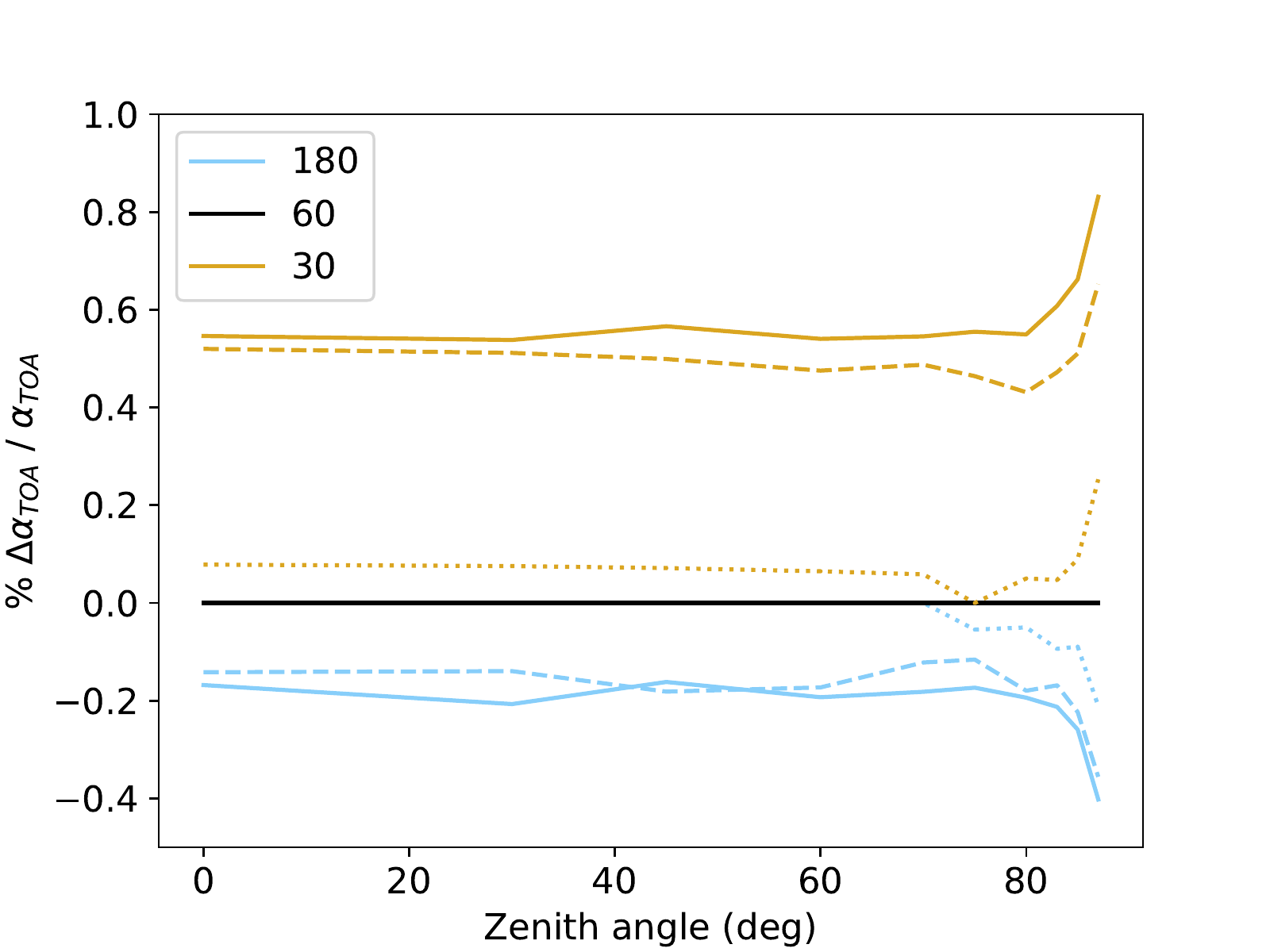}
\caption{The relative deviation (in \% points) from the TOA albedo of our reference model for Earth atmosphere (in black) of 30-layer (yellow lines) and 180-layer (light blue lines) models. On the left hand panel the dependence on temperature for $0^\text{o}$ zenith angle is shown. On the right hand panel the dependence on zenith angle for three temperatures  is shown: 240 K (solid), 280 K (dashed) and 360 K (dotted). \label{fig:atmogridTOA}}
\end{figure*}

First of all, we probed the stability of the EOS results with respect to variations of computational variables. We devised four tests to be conducted on the same specific atmosphere, namely the \textit{Saturated Modern Earth} one with 360 ppm of \cdiox, 1.8 ppm of \methane ~and a relative humidity (RH)  equal to 100$\%$ (see Table \ref{tab:models}). In each test the dry BOA pressure was set to 1 bar, the total BOA pressure was  varied according to the temperature-dependent contribution of water vapor and   we adopted a moist adiabat lapse rate topped by a 200\,K isothermal stratosphere (see Sec.~\ref{ss:earth_lapserate}). We then proceeded as follows.

In the first test we changed the layer density (i.e., the number of layers per order of magnitude of pressure) in the simulation. We ran \texttt{HELIOS} using 30, 40, 60, 120 and 180 atmospheric layers, for 7 points in temperature in the 240-360 K interval. The 60-layer model was taken as a reference. Since the contribution of the water vapor to the total BOA pressure is modest, the number of layers is independent of the surface temperature. The results can be seen in Fig.~\ref{fig:atmogrid}, left-hand panel. First, we note that reducing the density of layers has the effect of increasing the OLR, especially at intermediate (280-320 K) temperatures. For the 30-layer (5 layers per order of magnitude) the largest deviation from the reference model is found at 300 K. This deviation amounts to 3.9 W m$^{-2}$, corresponding to $+1.8\%$. For the lowest and highest point in temperature, the deviation reduces to less than 1 W m$^{-2}$, which corresponds to $+0.6\%$ and $+0.3\%$, respectively. 
Second, we note that increasing the density of layers produces smaller variations in the OLR. As an example, using 40 instead of 30 layers changes the OLR at 300 K by more than 2 W m$^{-2}$, while using 180 instead of 120 layers yields a difference of $\sim0.2$ W m$^{-2}$. Therefore, doubling or tripling the number of layers above the advised value of 10 layers per order of magnitude produces in general only marginal variations, in the order of 1 W m$^{-2}$ or $-0.3\%$ in the worst case.

In the second test (Fig.~\ref{fig:atmogrid}, right-hand panel), we have varied the pressure of the uppermost layer of the model, which for short we will call TOA pressure. Alongside the reference 0.1 Pa value, we have tried 1 and 10 Pa, either leaving constant the total number of layers (60, solid lines), or leaving constant the density of layers (10 per order of magnitude, dashed lines). First of all, increasing the TOA pressure when using a constant number of layers has the effect of decreasing the OLR, by $\sim 1$ W m$^{-2}$ in the worst case. This is somewhat counter-intuitive given that we are removing the outer part of the atmosphere from the calculations and therefore reducing the overall opacity of the column, but it can be explained in light of the results found in the previous paragraph. In fact, in this case we are indirectly increasing the density of layers in the lower part of atmosphere, much like we have done in the first test. On the other hand, keeping a constant density of layers nearly cancels these deviations, meaning that the choice of the TOA pressure is not particularly relevant, at least for an Earth-like chemical composition and within the tested temperature interval.

We were also interested in studying how the TOA albedo varied when the layer density of the model is changed. Therefore, as a third test, we calculated the shortwave reflected radiation when the atmosphere is illuminated by a blackbody at 5780 K, representing a G2-type star. For the surface, we have chosen a surface albedo $\boaa$ equal to 0.25 and independent of the wavelength. The results, in terms of $\Delta\toaa / \toaa$ percentage points, can be seen in Fig.~\ref{fig:atmogridTOA}. Differently from the OLR, whose peak deviation is located at mid-range temperatures, the maximum deviation of TOA albedo can be found at low temperatures, as can be seen in the right panel. The deviations, both for our 30-layer and for our 180-layer model, reduce to less than $0.1\%$ at higher temperatures. In the left panel the dependence of TOA albedo deviations from the ISR zenith angle z is shown for three different values of T$_\text{s}$: 240 K (solid), 280 K (dashed) and 360 K (dotted). We note that: (i) the deviations become larger for higher values of z and (ii), at all z the deviations reduce when temperature is increased. In any case, in the temperature range considered, the maximum deviation found is $<1\%$ for all values of z. As for the OLR, we stress that by increasing the number of layers above 10 per order of magnitude the improvements in accuracy become smaller and smaller.

In the fourth test we studied the TOA albedo variations in response to a change in the TOA pressure. In particular, we tested the case with a 10 Pa TOA pressure, first keeping constant the number of layers and then changing this number in such a way to have the same layer density as in the 0.1 Pa reference model. 
Increasing the TOA pressure while leaving constant the total number of layers produced a very slight ($\sim 0.1\%$) reduction of the TOA albedo at low temperatures and no reduction at high temperatures for a fixed z. Given that the differences are small, we do not include a plot for this test. Higher values of z presented a more pronounced deviation. When we adjusted the number of layers to have same layer density as in the reference model the aforementioned effects disappeared. 

\subsection{Changes in the atmospheric structure} \label{ss:atmopT}

\begin{figure*}[]
\includegraphics[width=0.33\textwidth]{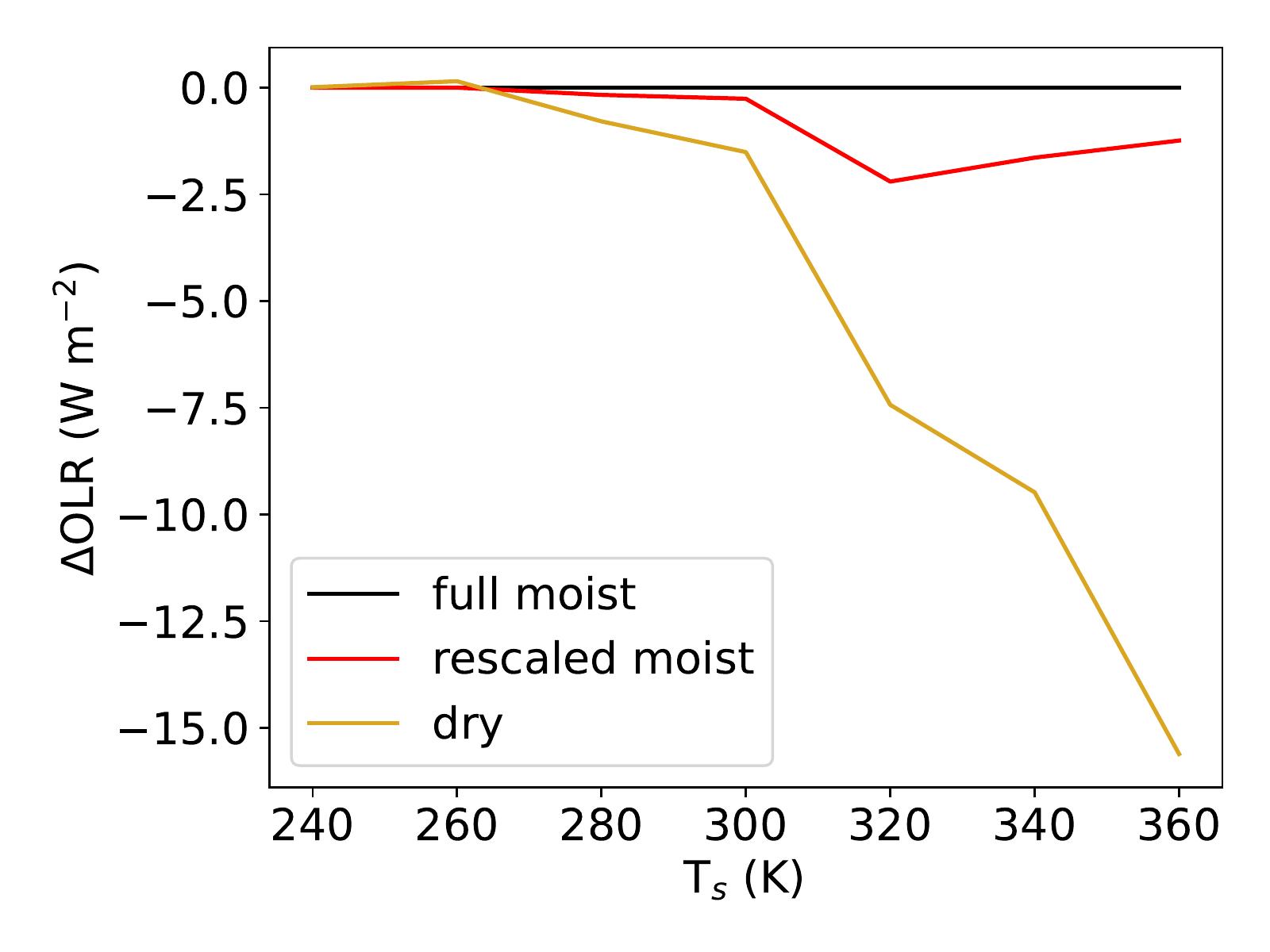}
\includegraphics[width=0.33\textwidth]{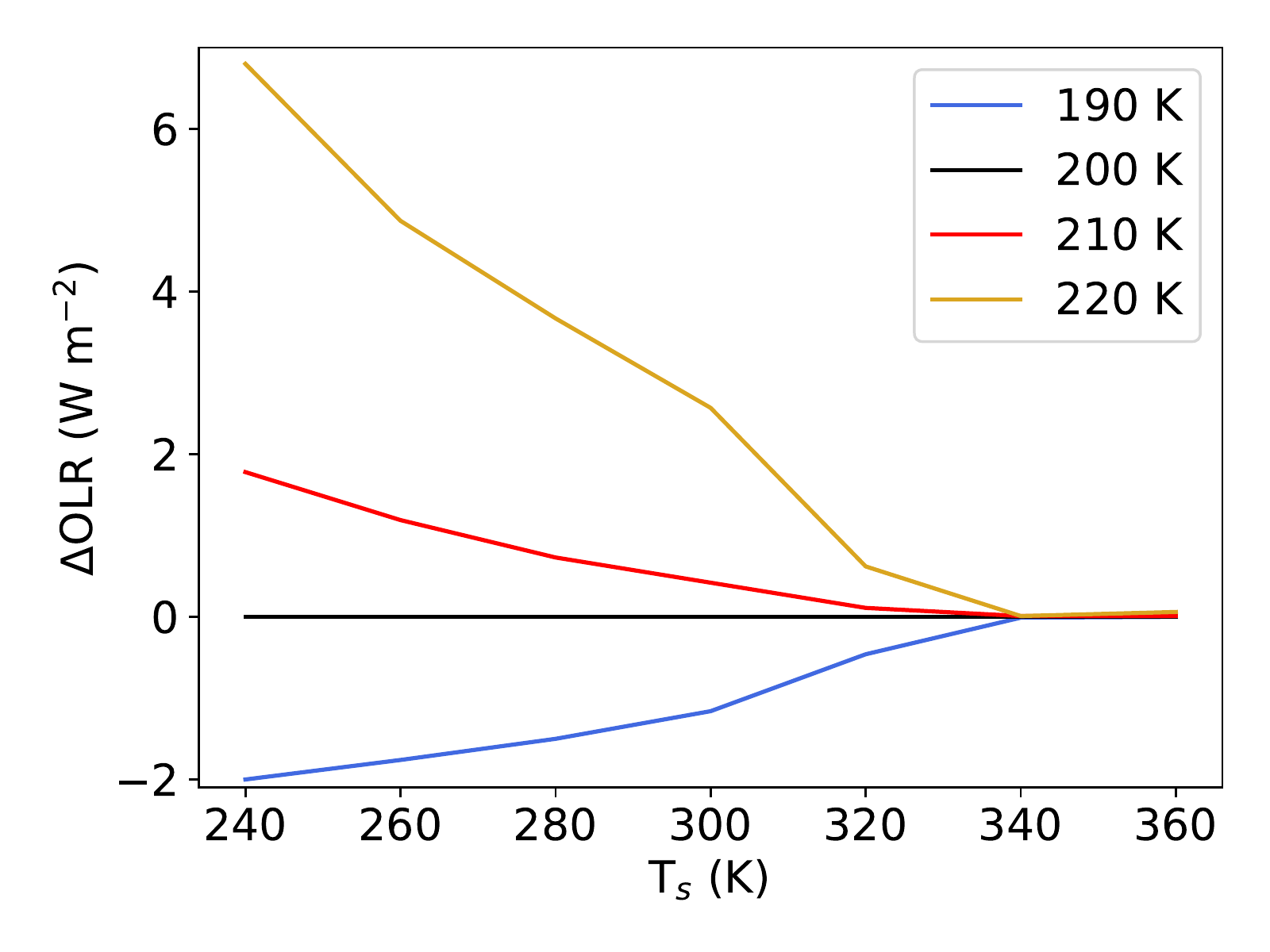}
\includegraphics[width=0.33\textwidth]{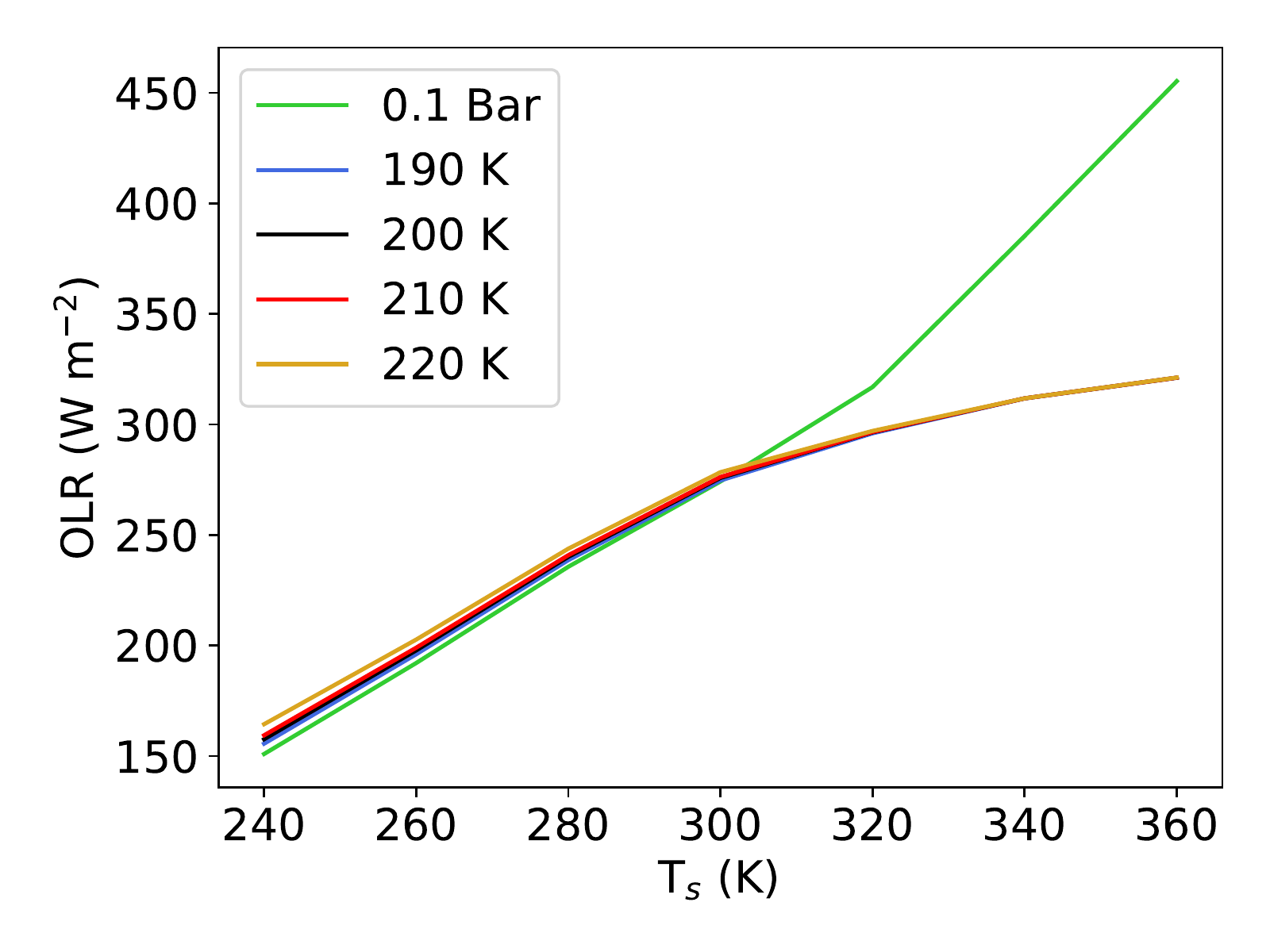}
\caption{
Comparison of the OLR predicted by different models of lapse rate (left panel) and tropopause prescriptions (central and right panels), as indicated in the legends. In the first two panels we display the absolute deviations, $\Delta$OLR, with respect to the reference model (full moist lapse rate and tropopause at 200\,K).
In the right panel we display the absolute value of the OLR. The 0.1 bar fixed-pressure tropopause case (in green) strongly deviates  from the fixed-temperature cases.
\label{fig:LRandTropo}}
\end{figure*}

\begin{figure*}[]
\plottwo{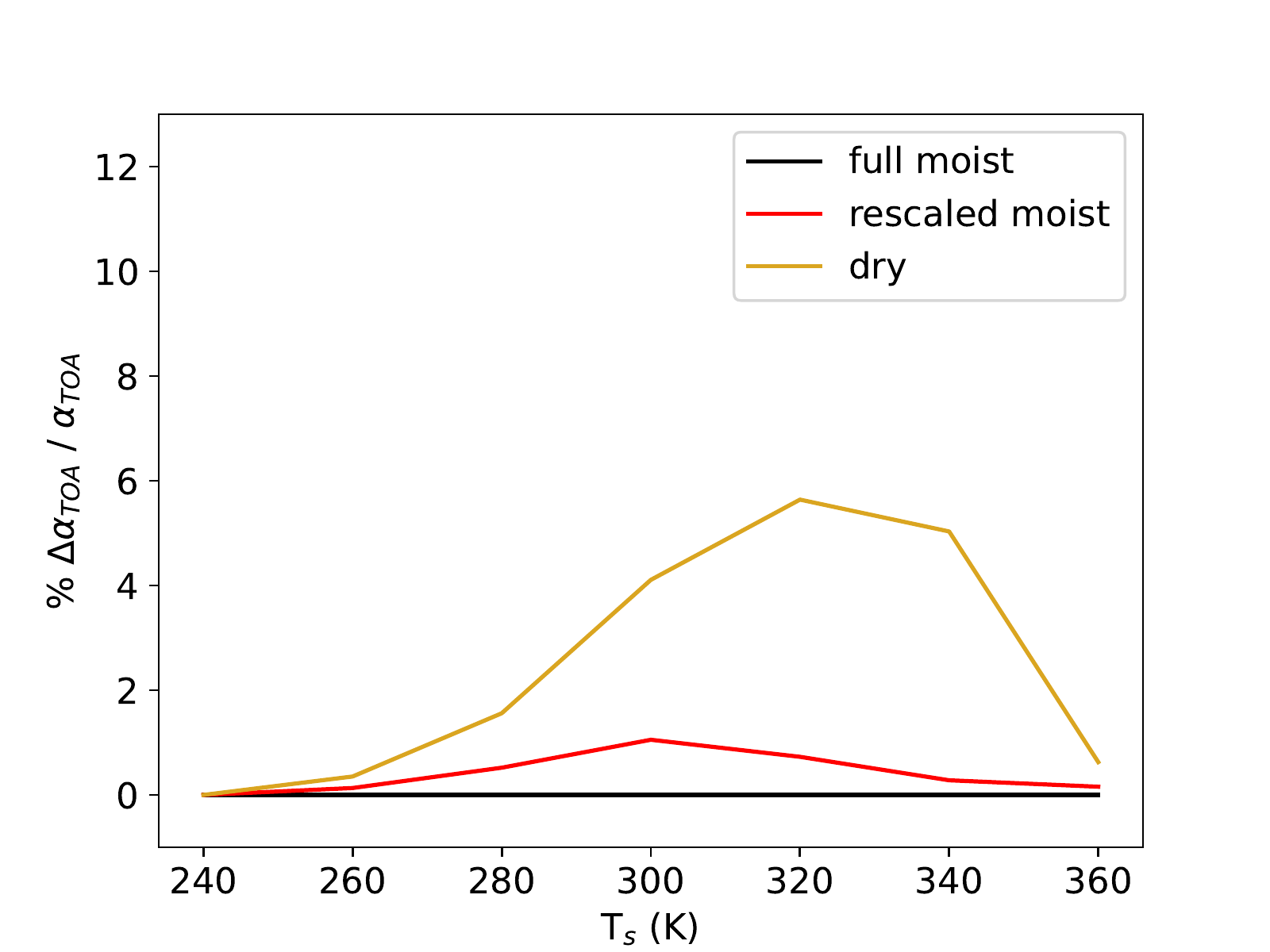}{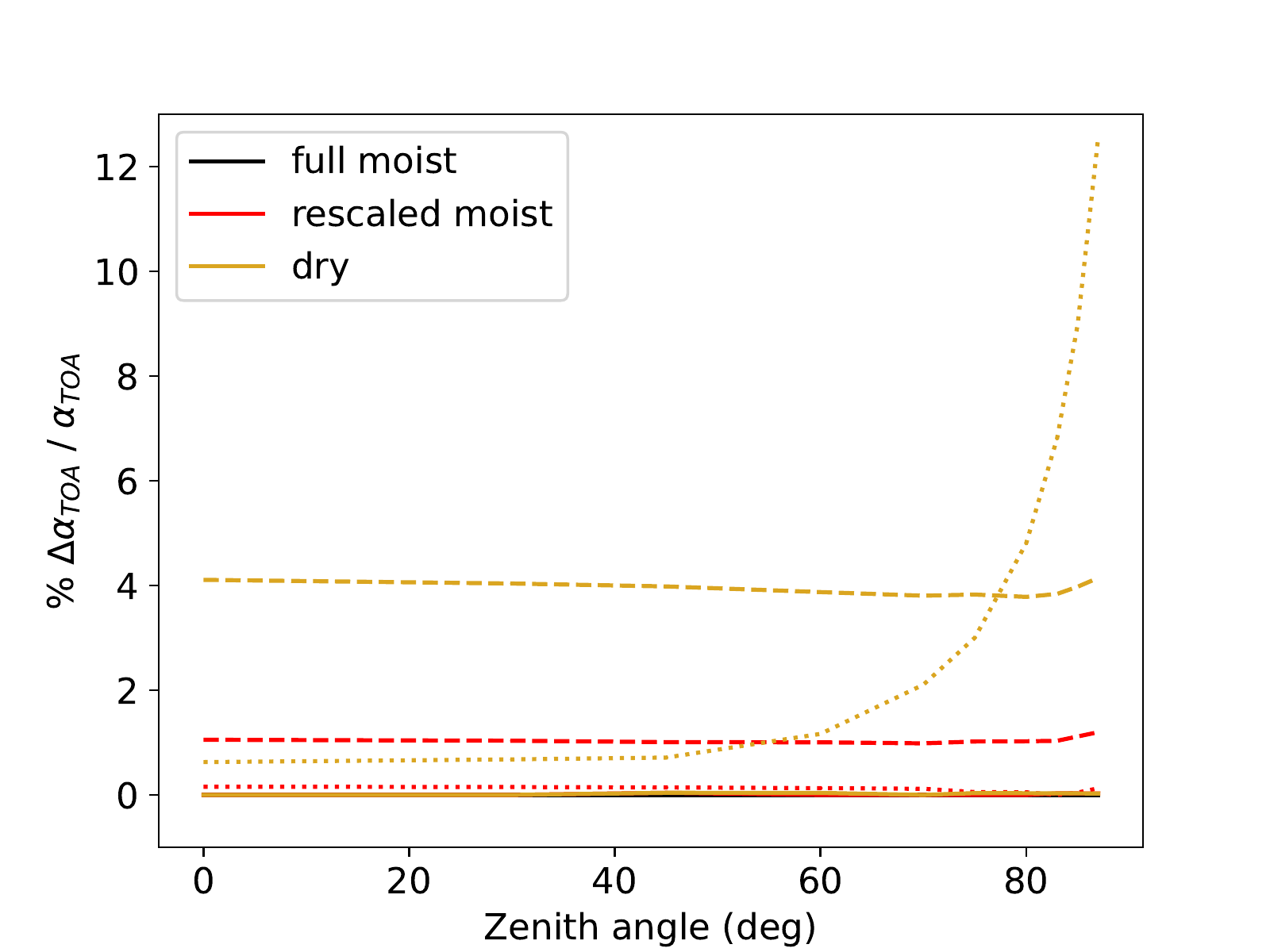}
\caption{The relative deviation (in \% points) from the TOA albedo of our reference model (ideal moist adiabatic lapse rate) of the dry adiabatic and the moist rescaled adiabatic lapse rate models. On the left-hand panel the dependence on temperature for $0^\text{o}$ zenith angle is shown. On the right-hand panel the dependence on zenith angle for three temperatures is shown: 240 K (solid), 280 K (dashed) and 360 K (dotted).
\label{fig:LRTOA}}
\end{figure*}

We then focused on the effects of changing prescriptions about the lapse rate and the tropopause position. In order to do so, we conducted two OLR tests and two TOA albedo tests on the same \textit{Saturated Modern Earth} atmosphere described in the previous subsection, using the reference model case values for the layer density (10 per order of magnitude) and the TOA pressure (0.1 Pa). 

In the first test, to study the impact on the OLR, we applied three different P-T atmospheric profiles, produced using a dry adiabatic, a moist adiabatic and the moist rescaled adiabatic lapse rate described in Sec.~\ref{ss:earth_lapserate}. 
Strictly speaking, the models with a moist rescaled lapse rate and with a dry adiabatic lapse rate are not self-consistent, because they refer to situations where \water ~is accounted for in the opacity calculations, but not in the determination of the vertical P-T structure of the atmosphere. On the other hand, the saturated moist lapse rate, which is commonly adopted throughout literature, despite being internally consistent, is not realistic since the average relative humidity of planetary atmospheres is lower than 100\%. 
With these caveats in mind, this test is useful to single out the influence of the lapse rate on the OLR, other factors being equal. The results are presented in Fig.~\ref{fig:LRandTropo}, left-hand panel. Changes in the lapse rates have maximum influence at high temperatures, where the atmosphere is richer in water vapor. At lower temperatures ($ T \lesssim $  300 K) even adopting a dry lapse rate, the OLR is reduced by less than 2 W m$^{-2}$ ($\sim 0.6\%$) with respect to our reference model (in black). Finally, our rescaled moist lapse rate produces results that are largely in line with those of a theoretical moist lapse rate, with peak differences of the order of $\sim 3$ W m$^{-2}$ for mid-to-high temperatures. Based on these results, in the standard application of our procedure we adopt the rescaled lapse rate rather than the moist or dry adiabat.

\begin{figure*}[t]
\plottwo{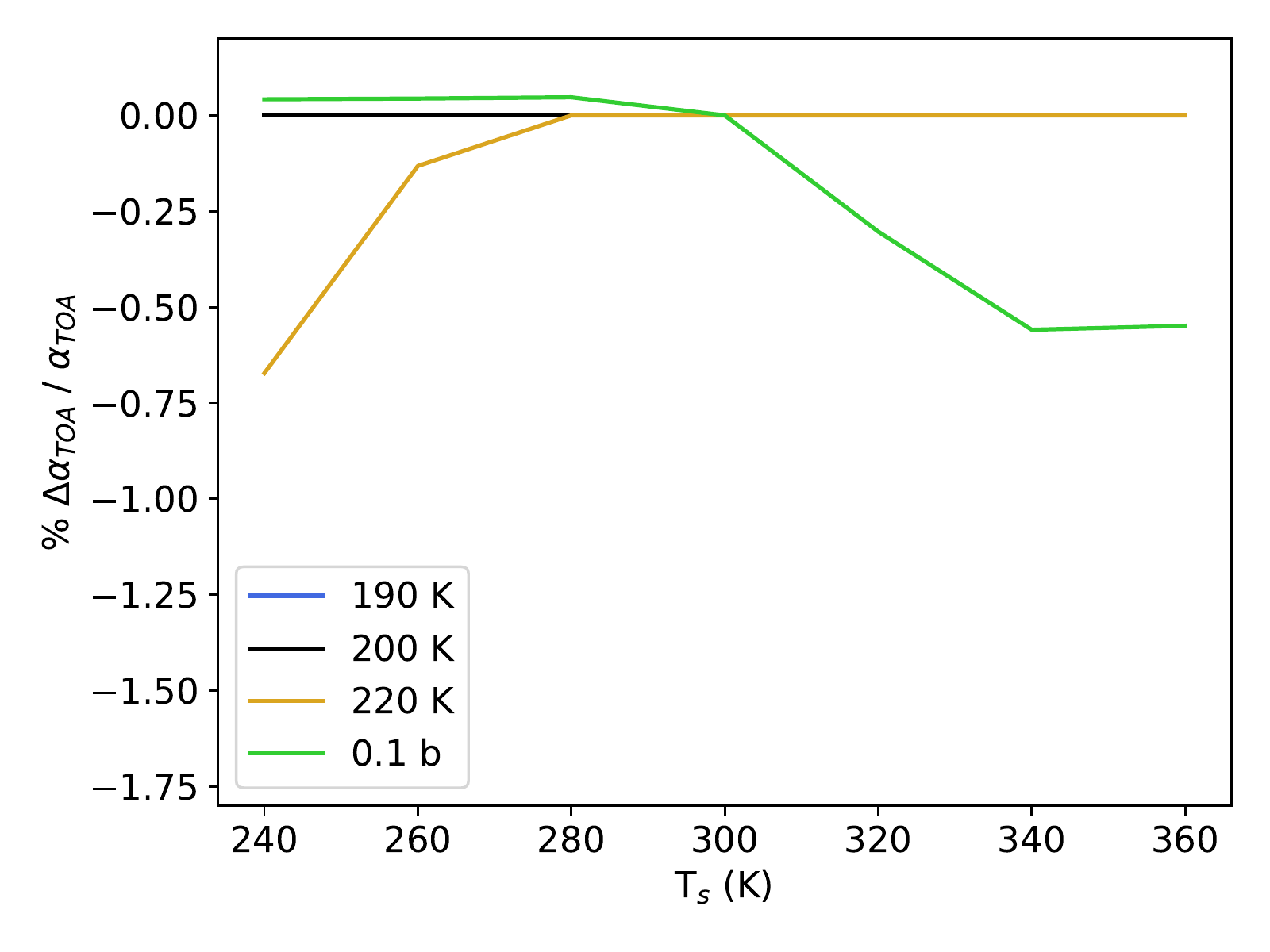}{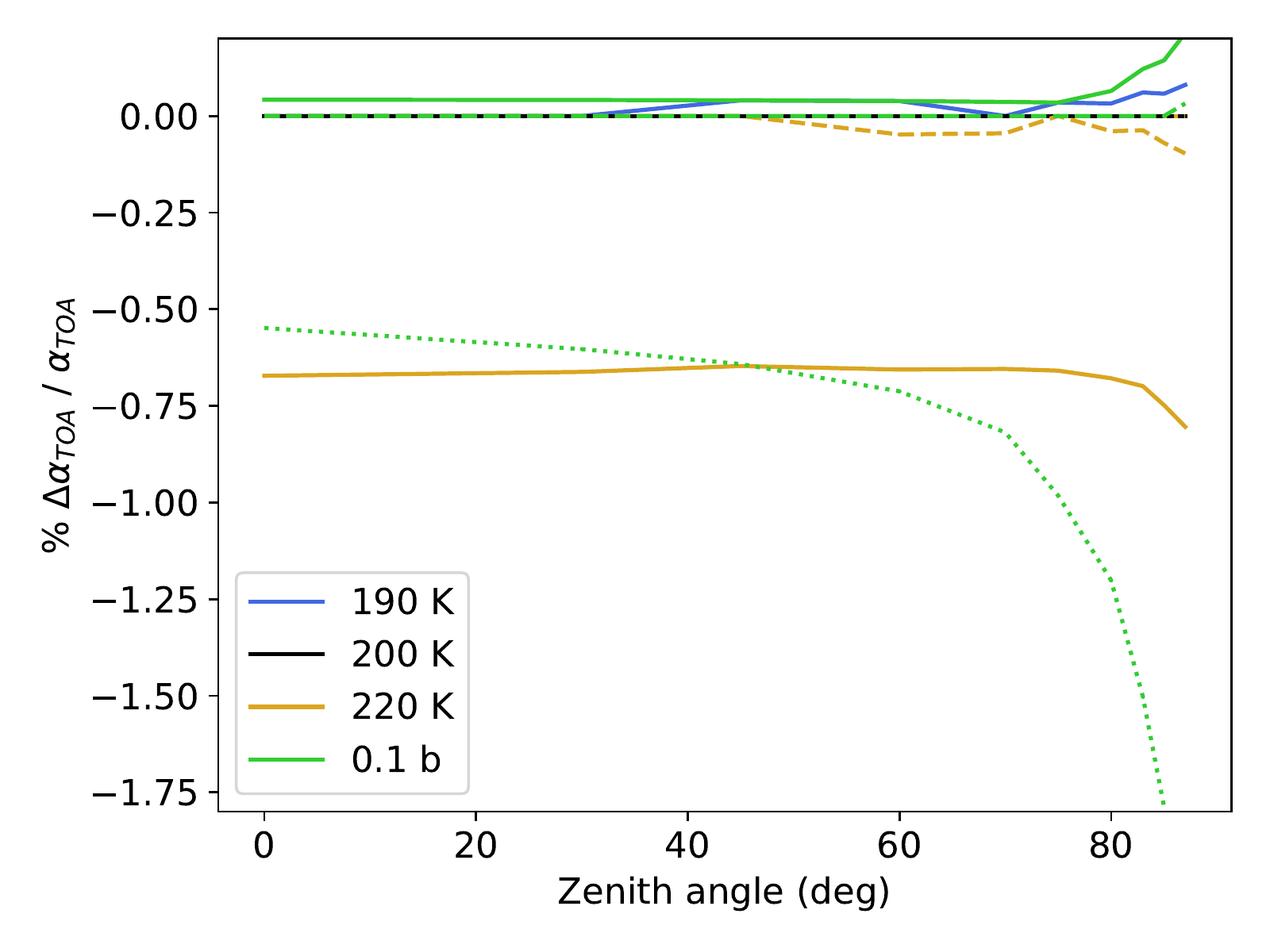}
\caption{The relative deviation (in \% points) from the TOA albedo of our reference model (200 K, in black) of 190 K (in blue) and 220 K (in yellow) models. On the left-hand panel is shown the dependence on temperature, for $0^\text{o}$ zenith angle. The gray line and the black line are coincident. On the right-hand panel is shown the dependence on zenith angle for three temperatures: 240 (solid), 280 (dashed) and 360 (dotted). Grey dashed and dotted lines and the black line are coincident.
\label{fig:TropoTOA}}
\end{figure*}

In the second test, we tried five different prescriptions for the tropopause position to study the effects on the OLR. Four of them are based on temperature and place the tropopause at the pressure corresponding to 190, 200, 210 or 220 K. The fifth prescription is taken from the work of \cite{robinson14}, who identified a common 0.1 bar position for the tropopauses of the Solar System planets. In this case, the tropopause is placed at the level corresponding to 0.1 bar. In all cases the overlying stratosphere has the same temperature of the tropopause. The results for the first four cases are shown in Fig.~\ref{fig:LRandTropo}, central panel. The maximum deviations from our reference model (namely the one with a 200 K tropopause) are found for the lowest temperature, while over 320 K the deviations becomes negligible (i.e. less than 1 W m$^{-2}$, or $0.3\%$). This behaviour is expected since, under these prescriptions, at high surface temperatures the tropopause is so high ($10^{-3}-10^{-4}$ bar) to be irrelevant on the final radiative balance of the atmosphere. On the other hand, the 0.1 bar tropopause prescription produces very different results, as it is possible to see in the right-hand panel of Fig.~\ref{fig:LRandTropo}. First of all, the deviations from our reference model are much larger both at low and at high temperatures, going from -6 W m$^{-2}$ at 240 K to +134 W m$^{-2}$ at 360 K. In particular, it is relatively small (some units of W m$^{-2}$) and negative in the 240-300 K range, and large (tens of W m$^{-2}$) and positive from 320 K onward. The second thing we note is the steeper OLR vs $\boat$ at high $T_s$ under the 0.1 bar prescription. This is contrary to what is expected when $T$ rises approaching the onset of the runaway greenhouse instability \citep{ingersoll69,nakajima92}: the troposphere becomes increasingly opaque to IR radiation, thus reducing the slope of the OLR as a function of $T_s$. Above some critical point, the OLR flattens and $T_s$ is allowed to increase freely up to the point in which the surface begins to radiate in the visible part of the spectrum \citep[see e.g. Fig. 4 of][]{kopparapu13a}. In our treatment this behaviour is reproduced when the tropopause is defined at constant temperature but not when it is defined at constant pressure. In this latter case the tropopause temperature becomes dependent on the surface temperature. Since the infrared photosphere of hot, \water-rich atmospheres is far above the 0.1 bar level, this prescription leads to a continual rise of the OLR with increasing surface temperature. If this situation were realistic, a runaway greenhouse would not be possible. However, the onset of the runaway greenhouse instability is predicted by a variety of models, including 3D GCMs, and is widely used to define the inner edge of the CHZ. The ability of reproducing such a feature is fundamental, therefore the fixed-pressure tropopause can be considered unfit to describe the climate and habitability of liquid-water bearing planets. In light of the above arguments we do not consider further the case of a tropopause at fixed pressure.

As in the previous subsection, we were interested also in studying the effects of the atmospheric structure changes on the TOA albedo. Therefore we did a third test in which we illuminated the three atmospheres described in the first test with a G2 star like blackbody. The wavelength-independent surface albedo was set to 0.25. The results, in term of percent deviation from our reference case, are shown in Fig.~\ref{fig:LRTOA}. On the left-hand panel, the dependence on temperature is shown. As it is possible to see, the deviation is negligible at low $\boat$ and low ($<1\%$) at very high $\boat$, while it is large at medium-to-high $\boat$ (300-340 K). On the right-hand panel the dependence of $\toaa$ on zenith angle is shown. As in most of the other cases presented in this paper, the deviation is larger for high zenith angles. This can be explained by the fact that, at high zenith angles, the optical path through the simulated atmosphere is longer, thus magnifying the differences between models.

Finally, in the fourth test we studied the variations in the TOA albedo when the tropopause prescriptions change. The cases analyzed are homogeneous to those of the second test. In Fig.~\ref{fig:TropoTOA}, left-hand panel, the percent deviations from our reference case (i.e., the one with a 200 K tropopause, black line) for the other tested prescriptions are presented as a function of surface temperature. We note three things. First, the effect using a fixed-temperature tropopause with a different temperature is maximum (0.7\%) at low surface temperatures and decreases rapidly at higher $\boat$ (yellow line). Second, using a 190 K tropopause yielded a deviation lower than 0.1 \% and therefore cannot be see on the plot. Third, when a 0.1 bar tropopause is chosen, the relation with temperature is inverted: the deviation is low at low temperatures and maximum at high temperatures. On the right-hand panel of Fig.~\ref{fig:TropoTOA} the percent deviation from the reference case is shown as a function of the zenith angle. As expected, deviations are negligible in the $0^o-75^o$ range and then increase,  but they remain small ($\sim$  0.1\% points). The only exception is the 0.1 bar troposphere case at 360 K (green line), which grows far more markedly to deviations larger than 2\% for high zenith angles.

\subsection{Changes in the atmospheric composition} \label{ss:compo}

\begin{figure*}[]
\plottwo{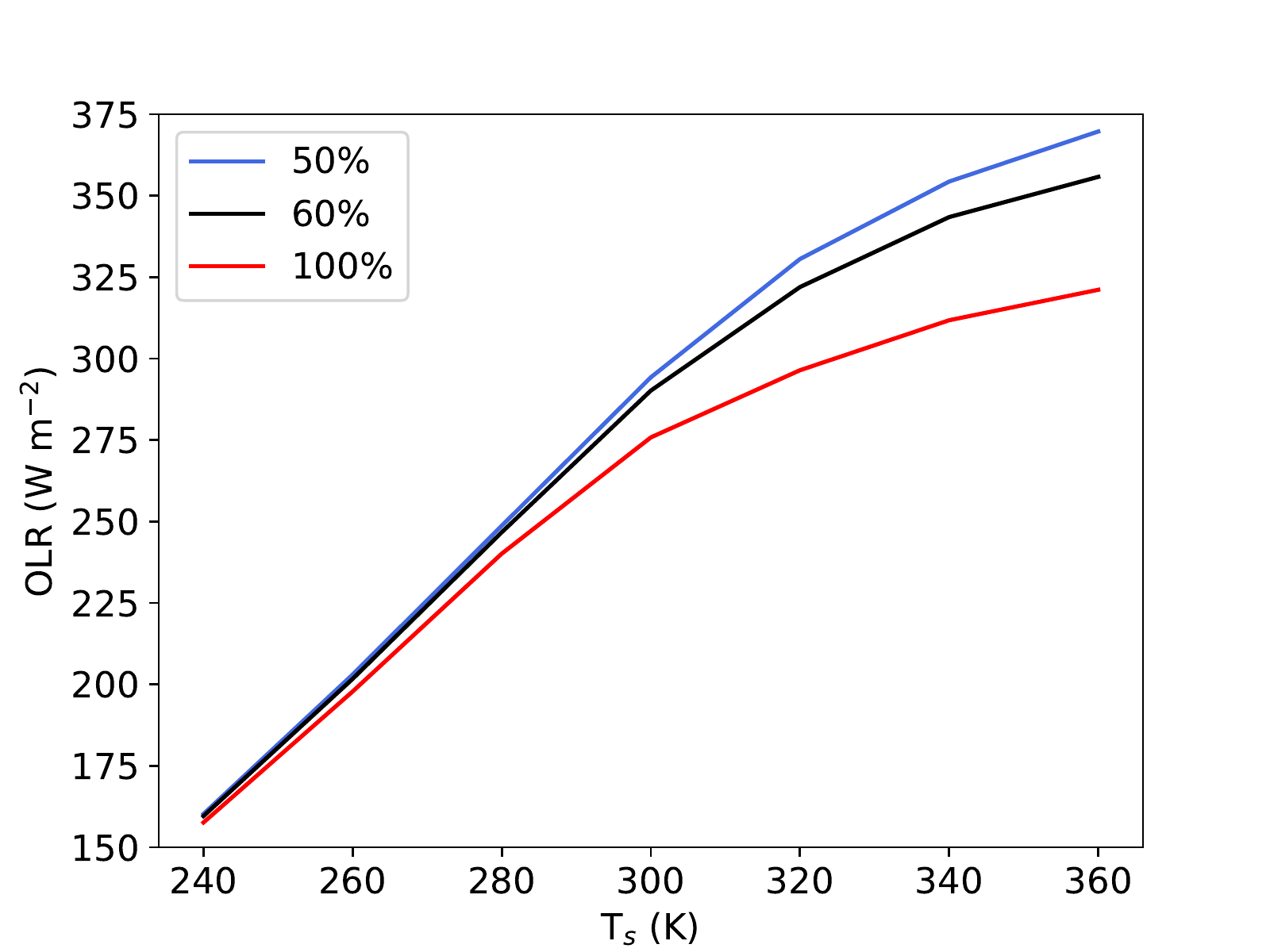}{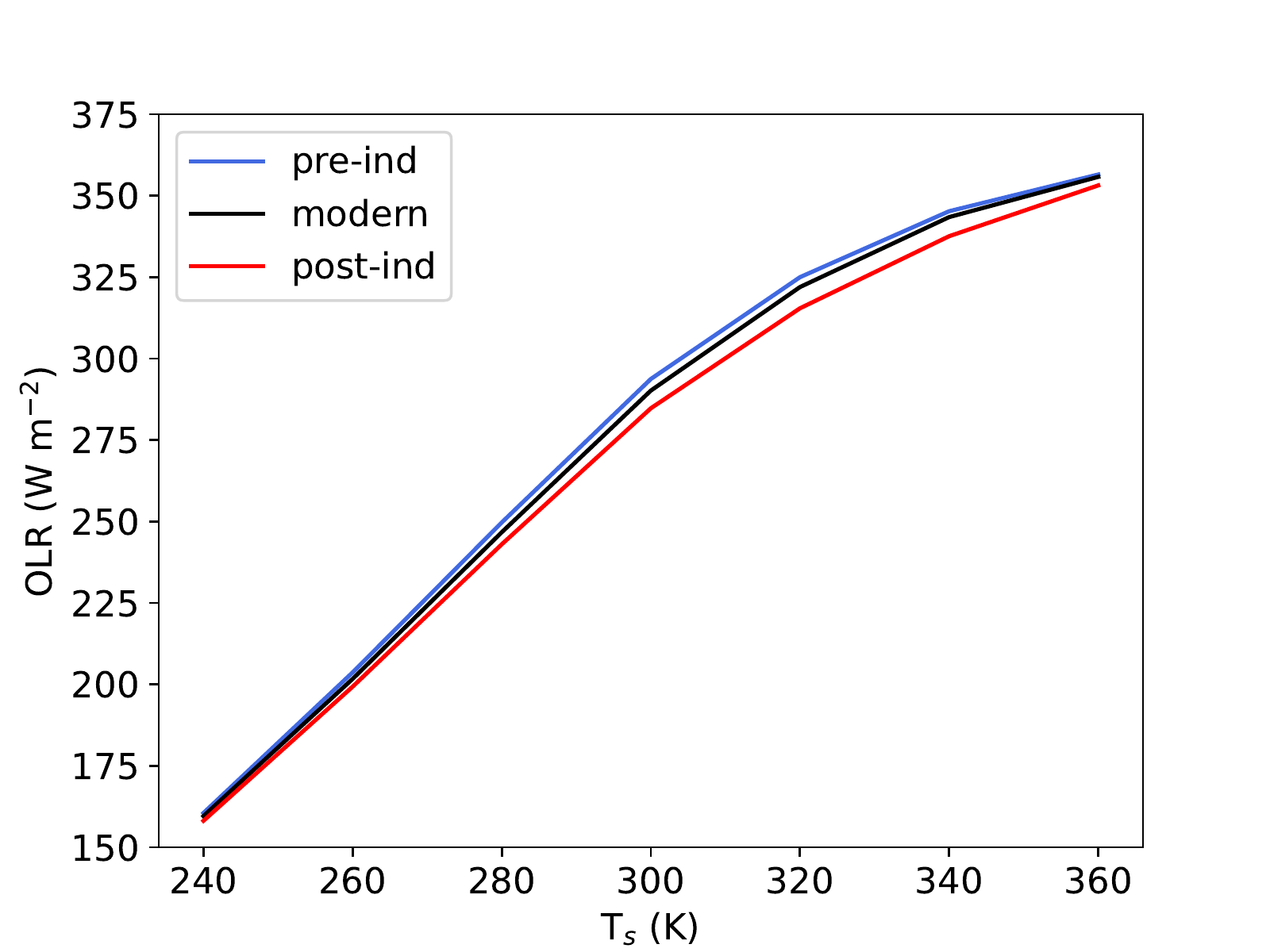}
\caption{The effects on the OLR of varying the relative humidity (left-hand panel) or the \cdiox ~and \methane ~dry molar fractions (right-hand panel). The reference model (in black) is the same in both panels and have RH=60\%, 360 ppm of \cdiox ~and 1.8 ppm of \methane. \label{fig:RHandCO2}}
\end{figure*}

\begin{figure*}[]
\plottwo{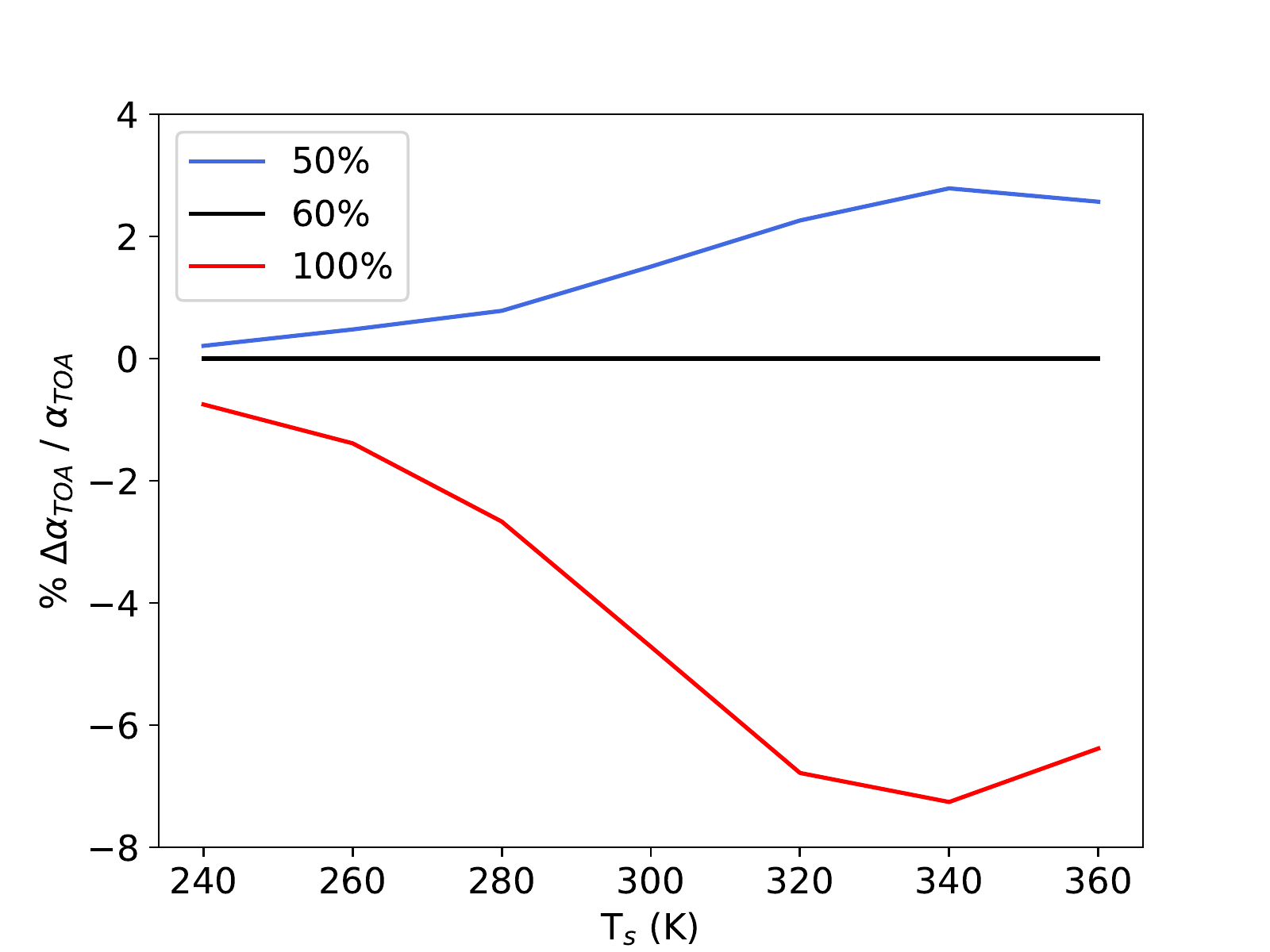}{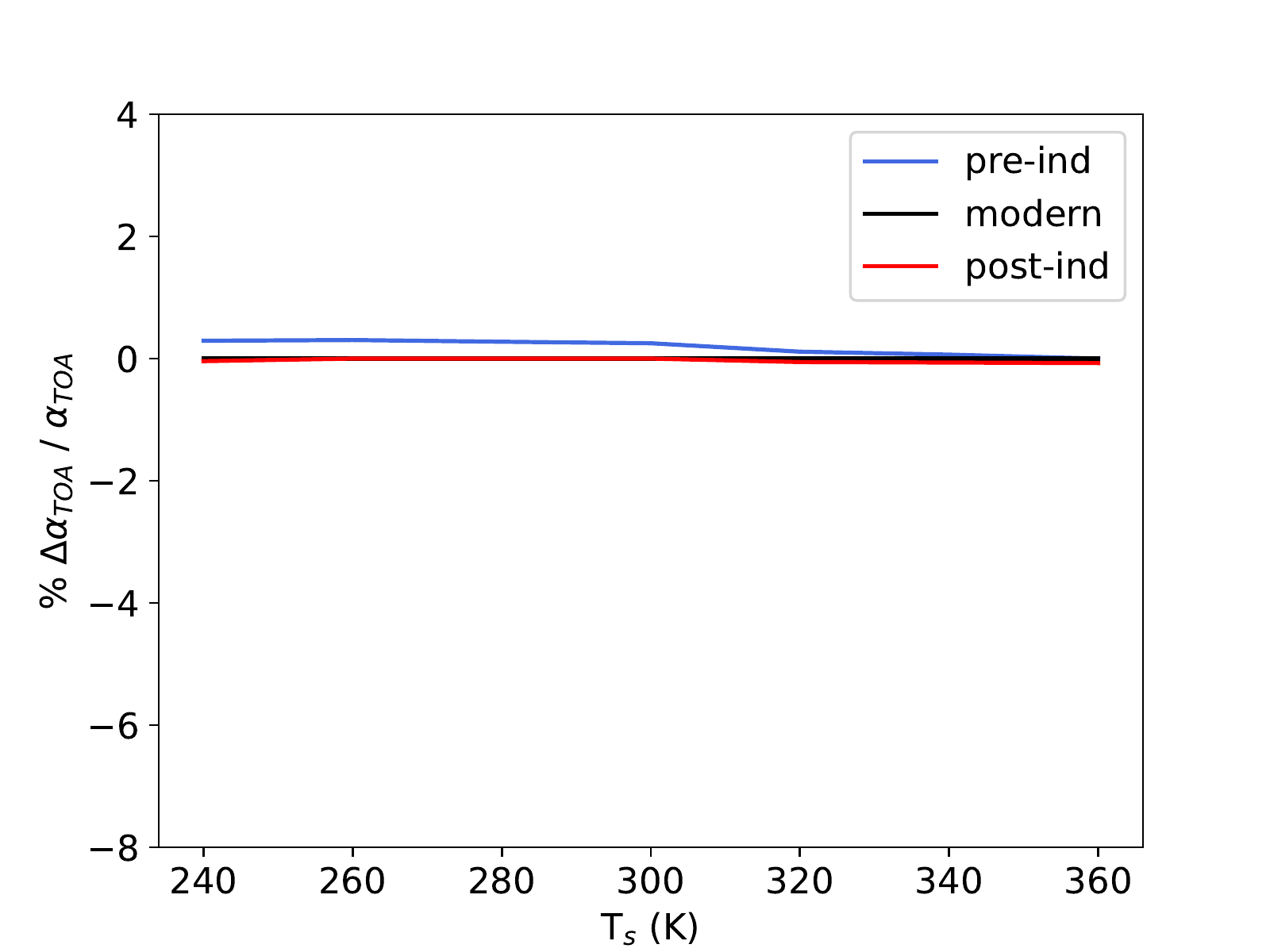}
\caption{The relative deviation (in \% points) from the TOA albedo of our chemical composition reference model (RH=60\%, pCO2=360 ppmv and pCH4=1.8 ppmv) for different values of relative humidity (left) and CO2/CH4 abundances. Both are plotted as a function of surface temperature. \label{fig:RHandCO2TOA}}
\end{figure*}

The most relevant effects on the OLR and TOA albedo come from the atmospheric composition. In order to quantify them, we have devised two tests for the OLR and two tests for the TOA albedo, illuminating the described atmospheres with a G2-like blackbody. 

In the first test, we varied the RH value of the atmosphere, testing three values: 50\%, 60\% and 100\%, which correspond to the \textit{Modern Earth 1}, \textit{Modern Earth 2} and \textit{Saturated Modern Earth} cases of Table \ref{tab:models}. The results are shown in Fig.~\ref{fig:RHandCO2}, left-hand panel. As expected, deviations start quite low at 240 K, with a difference between the 50\% and the 100\% case of 2.6 W m$^{-2}$ of 1.6\% and then grow quickly, becoming equal to 8.5 W m$^{-2}$ at 280 K, to 34.1 W m$^{-2}$ at 320 K and to 48.5 W m$^{-2}$ at 360 K. This behaviour is caused by the exponential increase of the water vapor content of the atmosphere as a function of temperature and its primary role as a greenhouse gas. Even a more modest increase, such as that present between the 50\% and the 60\% case is sufficient to produce a 0.6 W m$^{-2}$ OLR decrease at 240 K which grows to 1.9 W m$^{-2}$ at 280 K and to 13.9 W m$^{-2}$ at 360 K.

In the second test, we varied the \cdiox ~and \methane ~content of the atmosphere. We tested three cases: a \textit{Pre-industrial Earth} with 280 ppm of \cdiox ~and 0.6 ppm of \methane, a mid-1990 Earth (i.e.~the \textit{Modern Earth 2} case) with 360 ppm of \cdiox ~and 1.8 ppm of \methane\ and a scenario typically explored in climate simulations with 1152 ppm of \cdiox ~and no \methane, that here we call \textit{Post-industrial Earth}. In this last one, the \cdiox ~level is chosen as to simulate a 4 times increase of the pre-industrial \cdiox-equivalent (288 ppm) level, where \cdiox-equivalent stands for the quantity of \cdiox ~that produces the same OLR of a mixture of \cdiox ~plus other greenhouse gases (like \methane). The results are shown in the right panel of Fig.~7. The maximum deviations are found at 300 K for the pre-industrial case (+3.5 W m$^{-2}$) and at 320 K for the post-industrial case (-6.5 W m$^{-2}$). At high temperatures, the infrared opacity features of \cdiox ~and \methane ~are swamped by the stronger bands and continua of water vapor, which is the same throughout the models, thus reducing the relevance of those gases in the OLR computation and making the models to converge. On the other hand, at low temperatures, the deviations are lower in absolute terms but nearly proportional to the total OLR.

\begin{figure*}[]
\plottwo{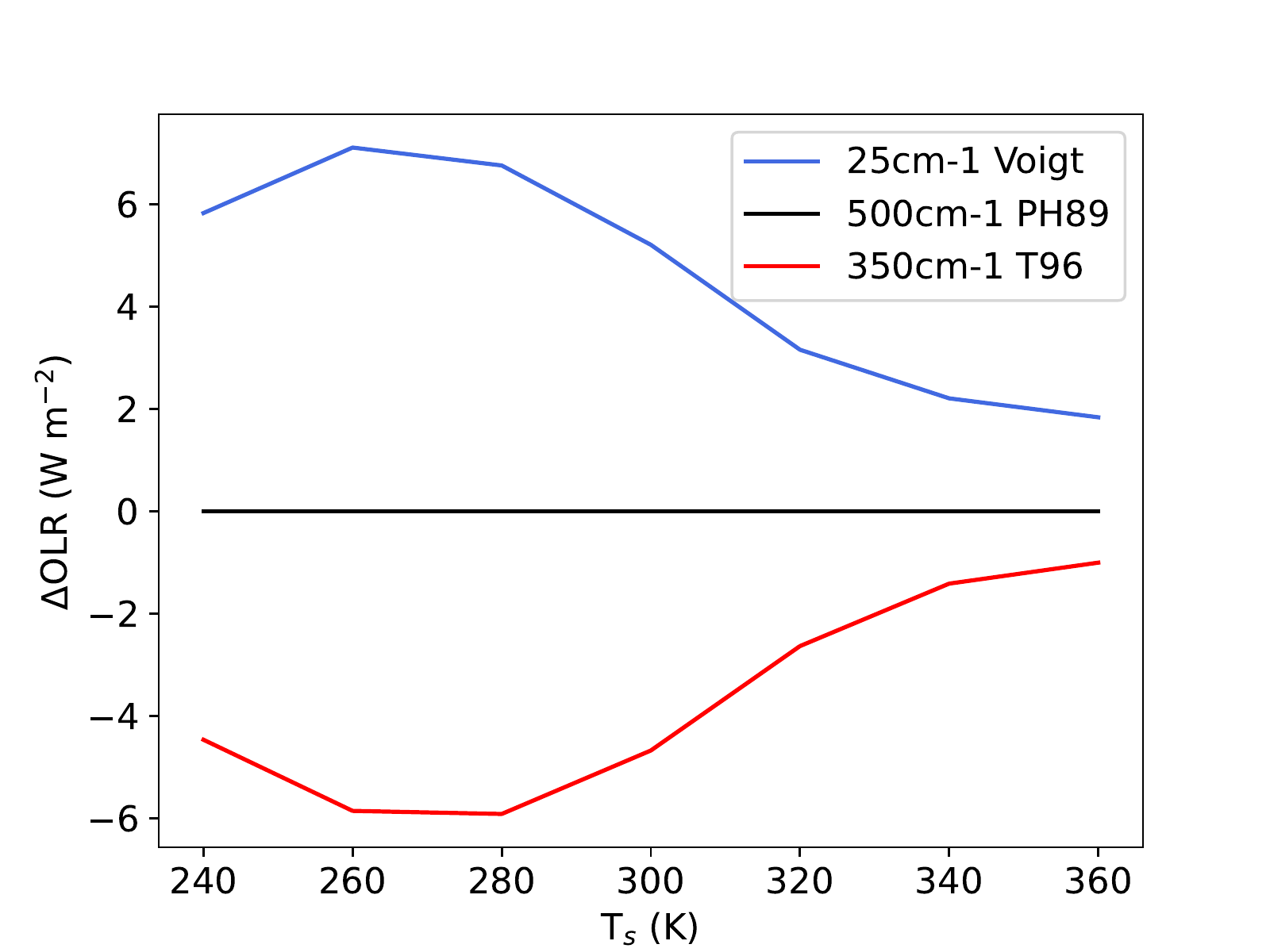}{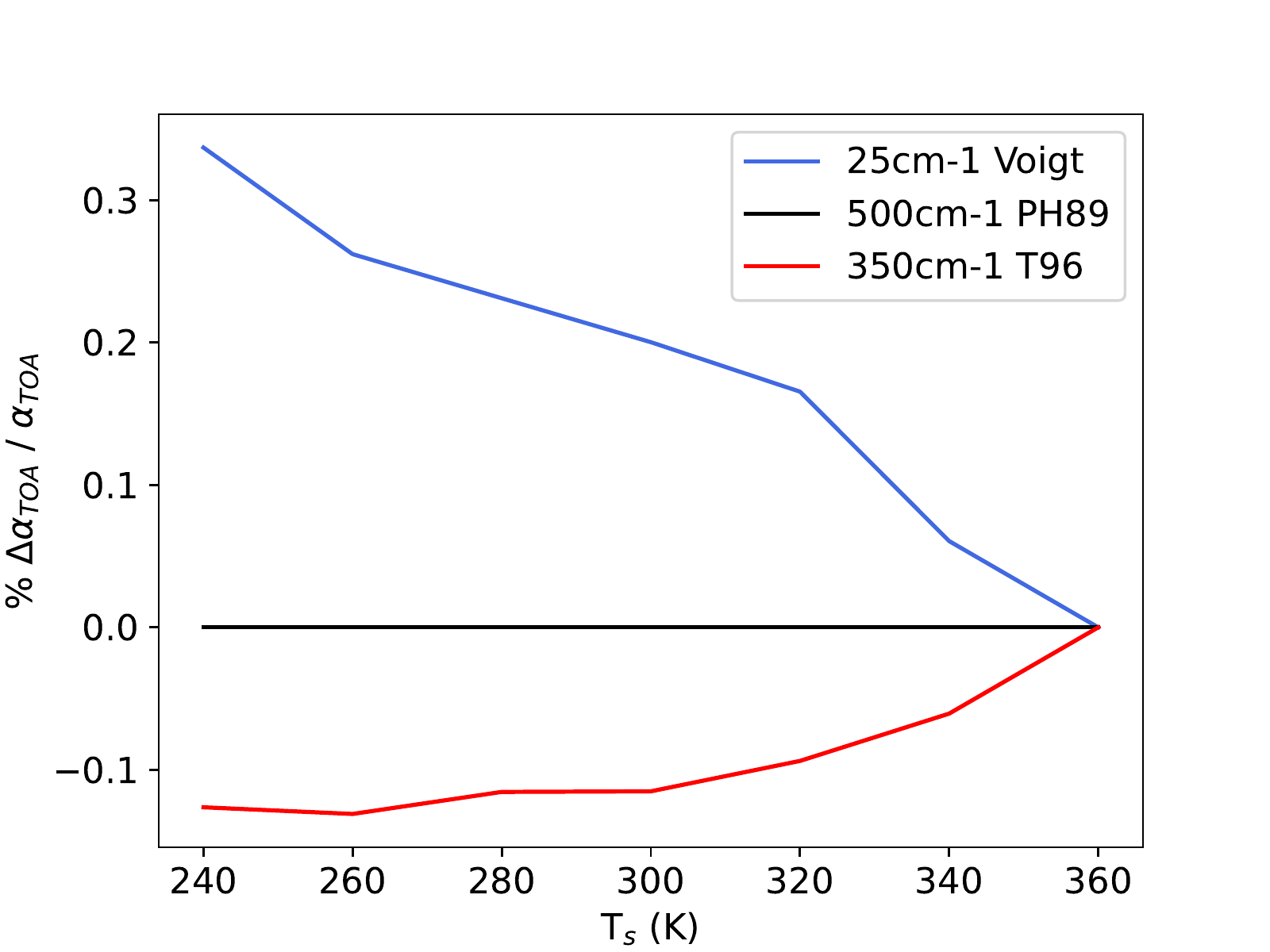}
\caption{The effects of varying \cdiox ~line profile prescriptions on the OLR (left-hand panel) and TOA albedo (right-hand panel). Both model cases use a 1 bar atmosphere of \cdiox ~with a water vapor humidity of 100\%. \label{fig:CO2Lines}}
\end{figure*}

In the third test, we studied the impact on the TOA albedo of changing the relative humidity. The results can be seen in Fig.~\ref{fig:RHandCO2TOA}, left-hand panel. Variations of TOA albedo are anti-correlated with the water vapor content in the atmosphere, mostly due to the larger absorption of water molecules in the near infrared. These differences are low at low temperatures (+0.2\% for the drier case, -0.7\% for the wetter case), increase at 340 K (+3.0\% for the drier case, -7.3\% for the wetter case) and then fall slightly at 360 K. We also tested how the TOA albedo changes as a function of the zenith angle (not shown)  and we found that, for higher angles, the deviations are nearly constant when $\boat$ is low and become smaller when $\boat$ is high. This is particularly evident for the 360 K case, where deviations reduce from +2.6\% to +1.4\% for the drier case, and from -6.4\% to -3.5\% for the wetter case. 

To complete our analysis, we studied how changes of \cdiox ~and \methane ~concentrations may affect the TOA albedo. As can be seen in Fig.~\ref{fig:RHandCO2TOA}, right-hand panel, the effects of such changes are very small, especially when compared to the effects of relative humidity changes. The maximum deviations are found for low $\boat$ and are equal to +0.3\% for the \textit{Pre-industrial Earth} case and to -0.1\% for the post-industrial one. Above 300 K, they become undetectable. When higher zenith angles are considered (not shown), two different things happen. For the Pre-industrial case, the positive difference at low $\boat$ (240 K run) is slightly amplified, changing from +0.3\% (at z=0$^o$) to +0.5\% (at z=87$^o$). This is caused entirely by the lower \cdiox~absorption, whose effect is amplified when the atmospheric optical path becomes larger at higher zenith angles. Instead, at higher $\boat$, this amplification becomes unnoticeable due to the dominant role of the water vapor opacity. On the other hand, for the \textit{Post-industrial Earth} case, the negative difference at low zenith angles become a positive difference at high zenith angles. Taking the 240 K run, the deviation evolves from -0.1\% to +0.4\%. For higher $\boat$, the effect becomes less strong, but still detectable. This increase of the TOA albedo can be explained by the fact that \cdiox ~is a very good scatterer compared to the main component of Earth-like atmospheres (\nitro), and for long optical paths the \cdiox~scattering may overcome the \cdiox~absorption. In these cases, \cdiox ~is a trace gas and its effect on the albedo is very small, but for \cdiox-dominated atmospheres the enhancement in the TOA albedo at high \cdiox ~partial pressures is so strong that it can overcome the increased greenhouse effect, thus cooling the surface. This is a well-known effect linked to the definition of the maximum greenhouse limit, which acts as outer limit for the circumstellar habitable zone (CHZ). A recent analysis of this phenomenon can be found in \cite{keles18}.

\subsection{Changes in the carbon dioxide line profiles} \label{ss:CO2wings_tests}

As described in Sec.~\ref{ss:CO2wings_theory}, \cdiox ~offers a special challenge in terms of absorption line profiles description. We tested three possible prescriptions for the \cdiox~lines: a Voigt profile truncated at 25 cm$^{-1}$, a PH89 profile truncated at 500 cm$^{-1}$ and a T96 profile truncated at 350 cm$^{-1}$. In all cases the CIA was considered. We carried out two tests to assess the effects of changing such prescriptions on the OLR and TOA albedo. In both tests we chose a 1-bar \cdiox\ atmosphere with a water vapor relative humidity of 100\% (that we call the \textit{Moist \cdiox-dominated} case). The vertical P-T profile was calculated as in Sec.~\ref{ss:co2_lapserate}. 

In the first test we investigated the OLR and the results are shown in Fig.~\ref{fig:CO2Lines}, left-hand panel. The truncated Voigt profile is less opaque, allowing for an increase in the OLR at low temperatures of 5.8-7.1 W m$^{-2}$, which is a 4.7\% increase with respect to the reference (PH89) model OLR. The T96 profile is instead more opaque causing a decrease of the OLR by 4.5-5.8 W m$^{-2}$ (i.e.\ $-4.2$\%). At higher $\boat$ the \water~absorption becomes increasingly important and the effects of different profiles decline to less than 0.5\%. We also removed the CIA absorption to assess its impact on the OLR. In all three cases, we find very similar results: between 240 and 300 K, the OLR increases by 5-7\% with respect to the same model with continuum absorption, while at higher $\boat$ the difference reduces to less than 1\% (not shown in figure).

In the second test we recorded the TOA albedo in the presence of an incoming radiation from our standard G2-star like blackbody.  The results are shown in Fig.~\ref{fig:CO2Lines}, right-hand panel. 
In this case, the relative differences between the different cases are basically negligible, peaking at 0.3\% for the Voigt profile at 240 K. The deviations produced by the T96 prescriptions are even less marked, peaking at 0.1\%. Finally, an increase in the zenith angle slightly reduces the deviations at all temperatures.

\section{Discussion}

\subsection{Comparison with previous RT calculations}
\label{ss:comparison}

We have compared the clear-sky OLR and TOA albedo predictions of EOS against those of other codes. We have considered an Earth-like atmosphere like in \citet[][hereafter Y16]{yang16} and a \cdiox-dominated atmosphere like in \citet[][hereafter H09]{halevy09}.

\subsubsection{Earth-like atmospheric models}

\begin{figure*}[]
\plottwo{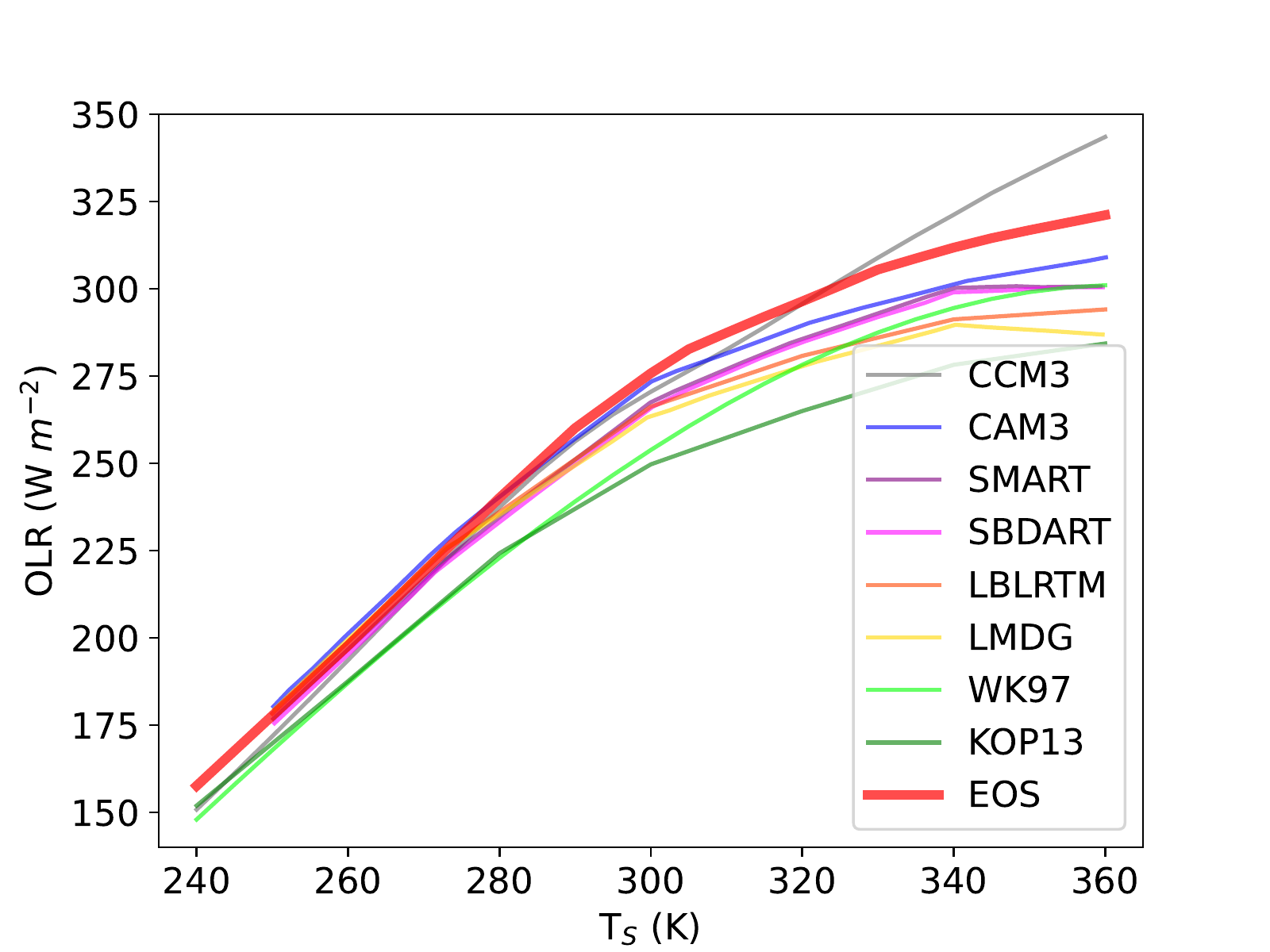}{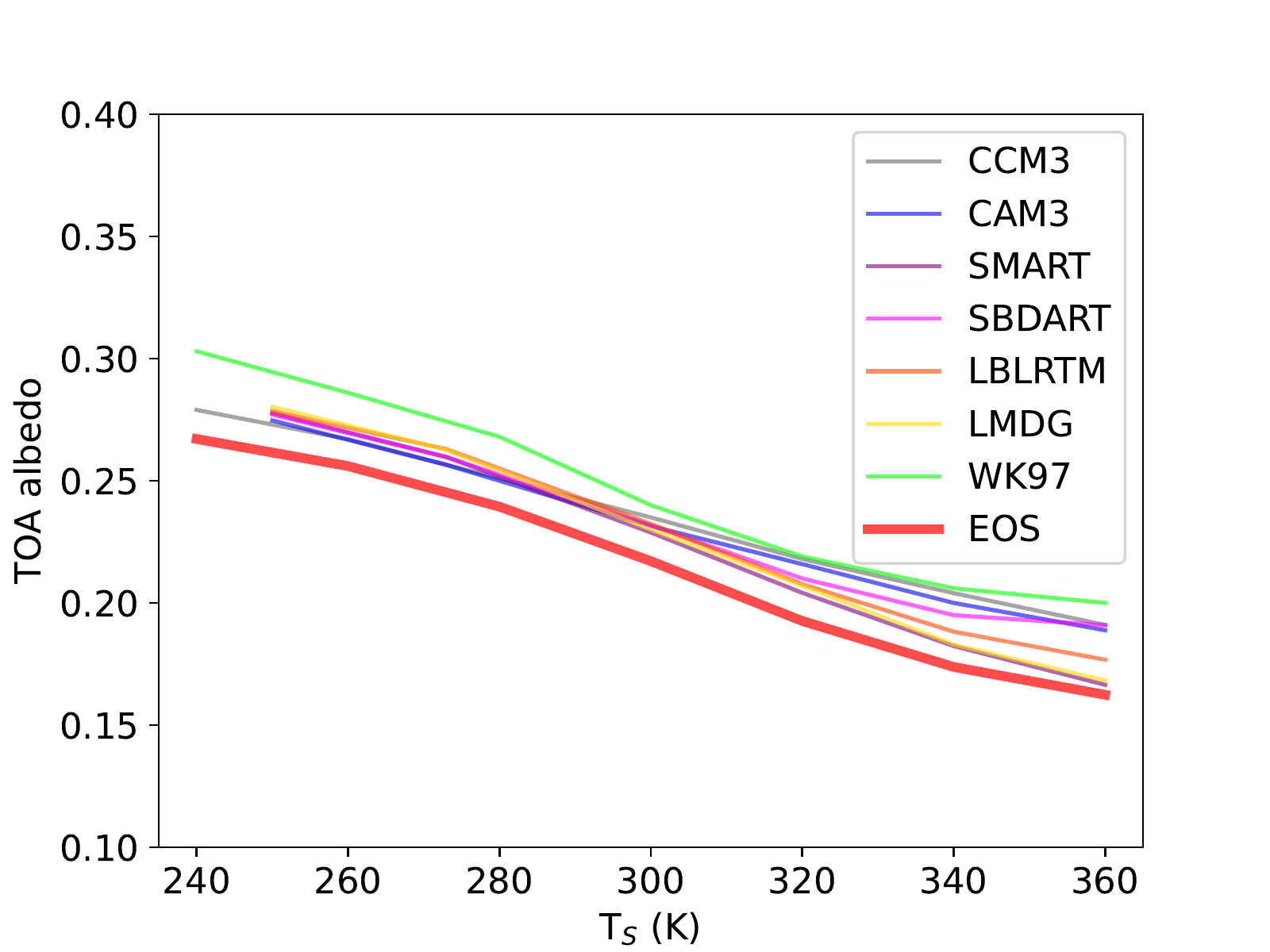}
\caption{The dependence on surface temperature of OLR (left-hand panel) and TOA albedo (right-hand panel) for 6 different RT codes, plus EOS (thick red line). Data relative to \texttt{CAM3, SMART, SBDART, LBLRTM} and \texttt{LMDG} has been taken from Y16, while data relative to \texttt{CCM3} has been produced by us. \label{fig:EarthComparison}}
\end{figure*}

\begin{figure*}[]
\includegraphics[width=0.33\textwidth]{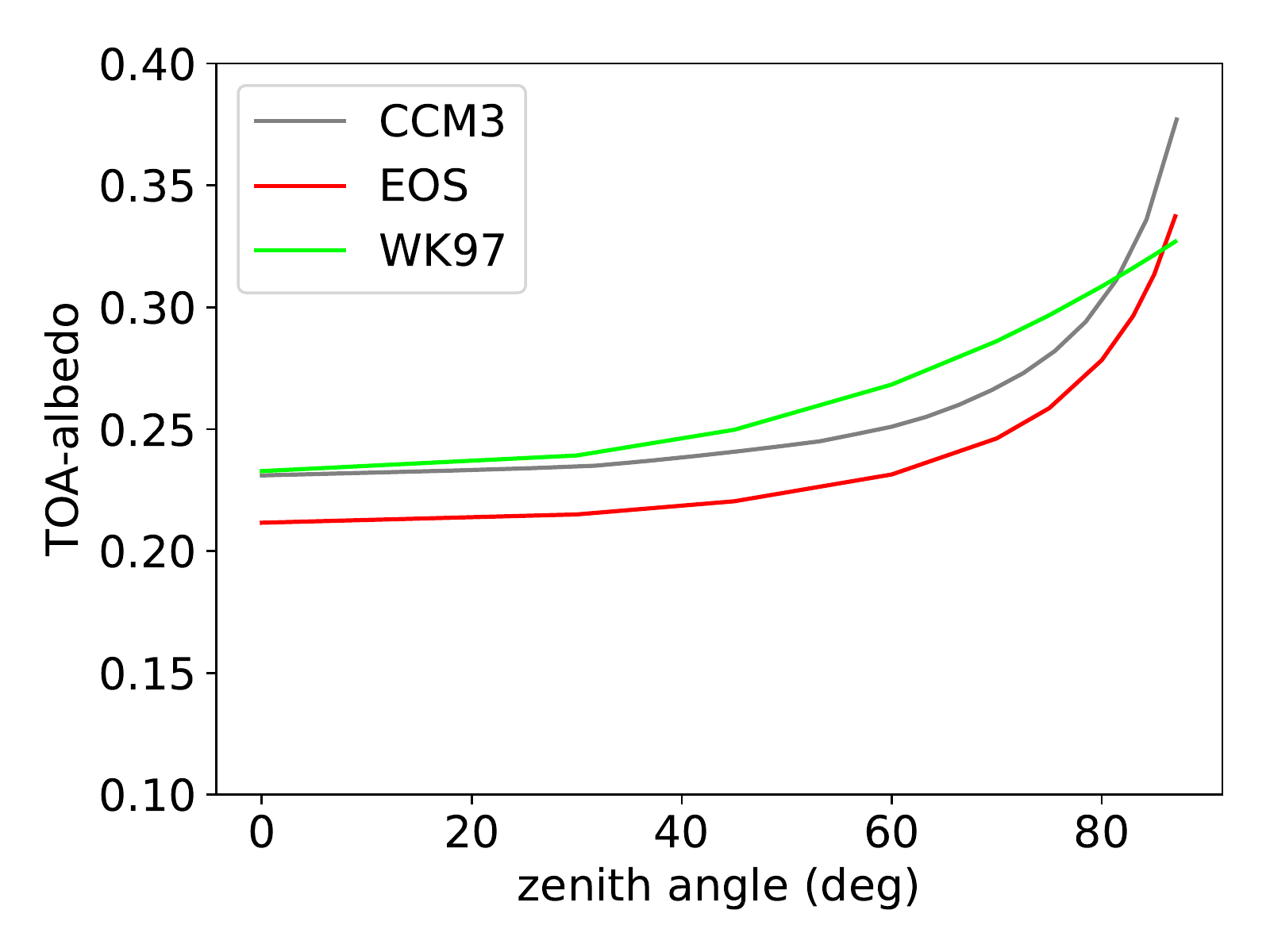}
\includegraphics[width=0.33\textwidth]{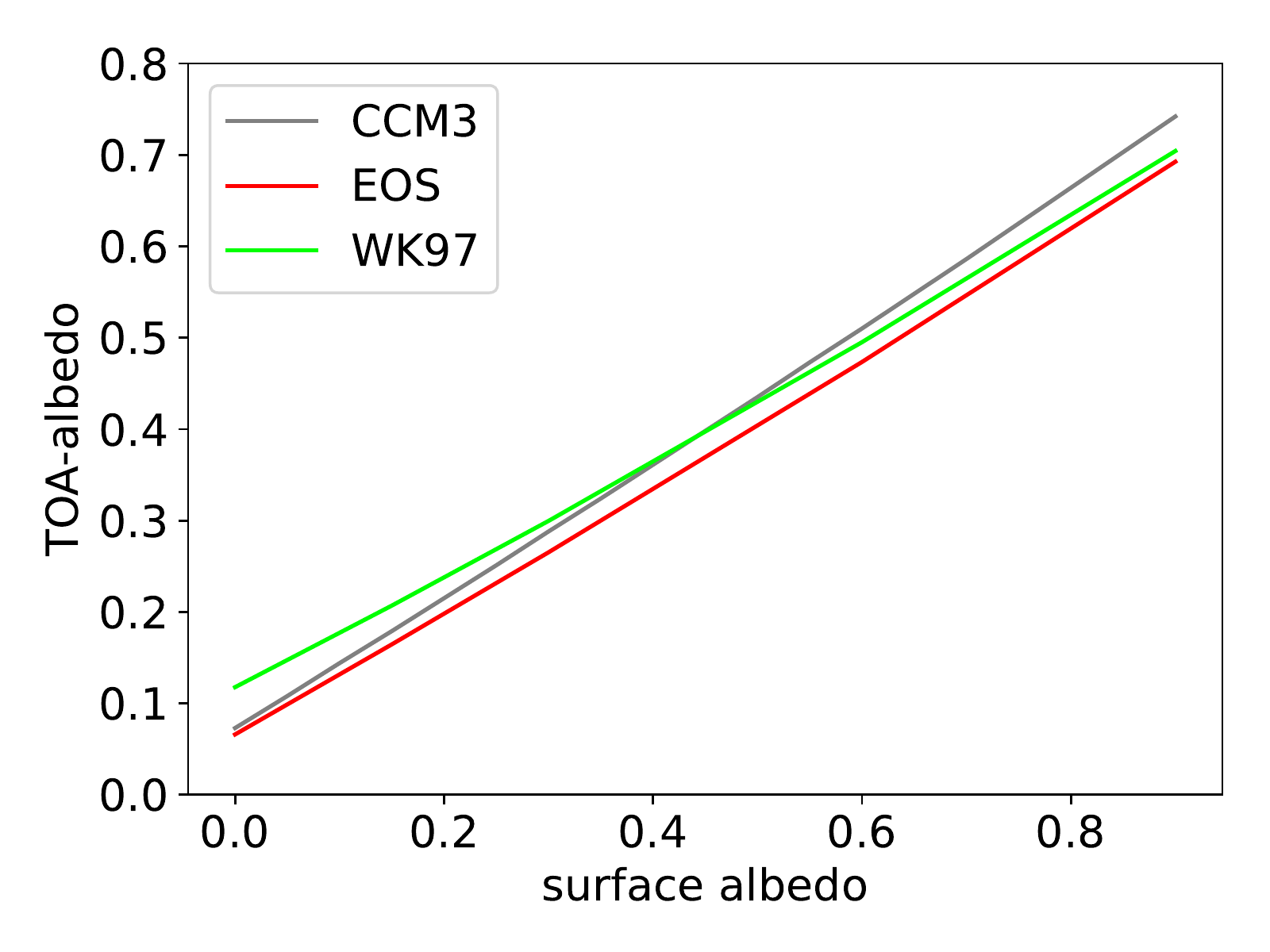}
\includegraphics[width=0.33\textwidth]{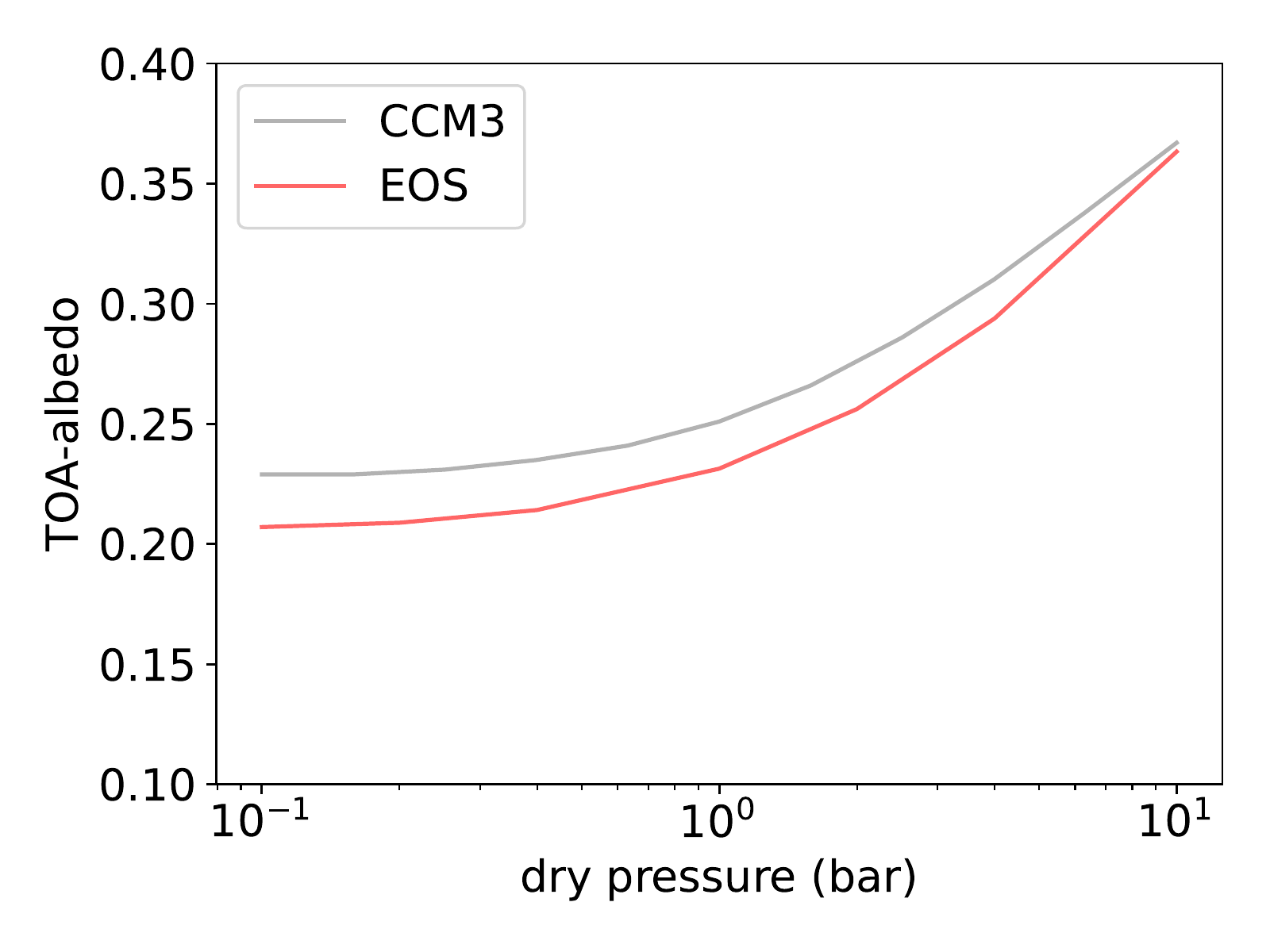}
\caption{The dependence of TOA albedo on the incident zenith angle (left panel), the surface albedo (central panel) and the surface dry pressure (right panel). On the latter one, given the limits of the WK97 polynomial formulation, we cannot compare their results with ours. \label{fig:TOAtrends}}
\end{figure*}

Concerning the Earth-like atmospheres we considered  five models discussed by Y16, namely \texttt{CAM3, SMART, SBDART, LBLRTM} and \texttt{LMDG}. In addition, we considered predictions of the \texttt{CCM3} \citep{kiehl98}  calculated by us as described in \cite{vladilo15}. Finally, we also included the results published by \cite{williams97}, hereafter WK97, and by KOP13.
The main characteristics of the models used as comparison tests can be summarized as follows.

\texttt{SMART} and \texttt{LBLRTM} are line-by-line RT models, \texttt{SBDART} is a high-resolution band model (with 369 spectral intervals), while \texttt{CAM3} and \texttt{LMDG} are band models with 19 and 36 intervals, respectively. All three band models were developed to study Earth-like planets around Sun-like stars, therefore their intervals are chosen so as to maximize the precision of calculation in this specific case. \texttt{CCM3} \citep{kiehl98} is another band model tuned for the study of Earth and Earth analogs. WK97 ran a large grid of calculations using a band model \citep{kasting86,kasting91} and fitted the resulting OLR and TOA albedo in polynomial form. KOP13 used a band model of their conception, based on the k-distribution opacity coefficients calculated by the \texttt{KSPECTRUM} tool. Their model uses 55 bands to calculate the OLR.
 
Each code adopts a different mix of prescriptions in terms of spectral line databases for optically active gases (\cdiox ~and \water), \water ~continuum absorption and Rayleigh scattering (see table 1 of Y16 and Sec.~2.1 of KOP13).

For \texttt{CAM3, SMART, SBDART, LBLRTM} and \texttt{LMDG} the atmosphere is composed by 1 bar of \nitro, 376 ~ppm of \cdiox ~and temperature dependent quantity of \water, calculated with a relative humidity of 100\%. In all these models the surface pressure was adjusted to consider the vapor pressure of \water, but the \cdiox ~mixing ratio was not  adjusted to account for the changes in the atmospheric gas mix. Therefore, high temperature points have a higher mass of \cdiox ~in the atmospheric column. 
However, Y16 estimate that the effect of this error on the OLR should be about 2.1 W m$^{-2}$ in the worst case (the 360 K point), which is $\sim 0.7$\% of the total OLR averaged through those models. The TOA pressure was set to 10 Pa and the number of atmospheric layers to 301, with the exception of \texttt{LBLRTM} which has 150 longwave layers and 75 shortwave layers and SMART which has 75 layers. The atmospheric vertical structure is the same as described in Sec.~\ref{sec:model} and shown in Fig.~\ref{fig:lapserates}. 

For \texttt{CCM3} the atmosphere is composed of 380 ppm of \cdiox, 1.8 ppm of \methane ~and variable \water ~with a relative humidity of 100\%. Adopting a value of 28 for the \cdiox-equivalent of methane gas \citep{ipcc21} gives a total amount of \cdiox-equivalent for the \texttt{CCM3} model of 430 ppm.

WK97 used an atmosphere composed of 1 bar of \nitro ~and saturated with \water. The partial pressure of \cdiox ~is an input of the polynomial that we have chosen in such a way to obtain a volume mixing ratio of 376 ppm, as in the models described by Y16. Regarding the vertical P-T structure, WK97 adopted a fixed tropospheric lapse rate equal to 6.5 K km$^{-1}$ and a temperature for the tropopause that is dependent on $T_s$ (see their equation A19). As a result, their model underestimates the OLR with respect to the other models at low T$_s$ and slightly overestimates it at high T$_s$. On the basis of our previous results (see Fig.~\ref{fig:LRandTropo}, central panel), OLR differences should be of the order of units of W m$^{-2}$, while TOA albedo differences should be negligible.

For the calculation of the inner edge of the habitable zone, KOP13 used an atmosphere composed of 1 bar of \nitro, 330 ppmv of \cdiox ~and saturated with \water. Their model atmosphere is divided into 101 layers. The vertical P-T structure is given by a moist pseudoadiabat, as described in the appendix A of \cite{kasting88}, overlain by a 200 K isothermal stratosphere.


Finally, for EOS we adopted the reference \textit{Saturated Modern Earth} atmosphere described in Sec.~\ref{ss:earth_lapserate}: 360 ppm of \cdiox, 1.8 ppm of \methane ~and 100\% relative humidity. 
The minor differences of trace greenhouse gas concentrations between different models are expected to produce completely negligible effects in the TOA albedo and very small differences in the OLR. In particular, for the latter we estimated a deviation of $\sim 0.5$ W m$^{-2}$, given that a doubling of \cdiox ~concentration produces a forcing of 4 W m$^{-2}$ \citep{collins06}.


Regarding the OLR, the results are shown in Fig.~\ref{fig:EarthComparison}, left-hand panel. As can be seen, for low temperatures (less than 280 K) EOS gives values that are perfectly in line with those obtained by other codes. At 250 K (lowest temperature point for the data taken from Y16), the highest OLR is given by \texttt{CAM3} (180.3 W m$^{-2}$), the lowest by \texttt{CCM3} (171.8 W m$^{-2}$), while EOS gives an intermediate result (177.7 W m$^{-2}$). Differences between all models start to arise when T$_s$ increases above 280\,K, presumably as a result of the differences in the treatment of the water vapor absorption. At very high T$_s$ the slope of the OLR vs T$_s$ flattens in most models, including EOS, a behaviour that is expected at the onset of the runaway greenhouse instability (see previous discussion in Sec.~\ref{ss:atmopT}). Only \texttt{CCM3} shows a steady rise of the OLR with increasing T$_s$, probably due to its outdated prescriptions. 
In some of the models tested by KOP13, precisely in those lacking the prescriptions of the \water ~continuum absorption, the flattening of the slope takes place at temperatures that are higher than in their reference models.
Clearly, the treatment of the effects of water vapor at high $\boat$ is extremely difficult and model dependent. EOS, with its updated prescriptions, is able to capture these effects at high $\boat$, but with an OLR that is higher than that predicted by the codes compared in Y16.
A satisfactory explanation of this difference would require a thorough analysis of each of the RT codes described in that paper, which is something beyond the scope of this work. However, the adoption of EOS predictions in \texttt{ESTM} has allowed us to reproduce a variety of zonally-averaged observables of the current Earth climate while keeping the other \texttt{ESTM} parameters well in the range of Earth observational data \citep[][in preparation]{biasiotti21}. This encourages us to believe  that the quantities calculated by EOS are physically sound. 

Regarding the TOA albedo under a G2-like blackbody insolation, the results are shown in Fig.~\ref{fig:EarthComparison}, right-hand panel. All models, including EOS, predict a decreasing of TOA albedo as T$_s$ increases, but the EOS albedos are slightly lower than those predicted by the other codes, with typical differences in the order of $\simeq -0.02$, and $-0.04$ in the worst cases. We have not included the results of KOP13 because they used a different value for the surface albedo (0.32). Using \texttt{CCM3} and the WK97 polynomials, we were able to compare also how the EOS-calculated albedos as a function of the zenith angle, the surface albedo and the surface dry pressure. These further results are shown in Fig.~\ref{fig:TOAtrends}. As it is possible to see, all the trends are reproduced. The small offset can be in part explained by two known differences between EOS and the other codes, namely (i) the fact that EOS corrects the zenith distance for the sphericity of the planet (for both Fig.~\ref{fig:EarthComparison}, right panel and Fig.~\ref{fig:TOAtrends}, left panel) and (ii) the usage of a slightly hotter blackbody for the results of Y16 and \texttt{CCM3} (for Fig.~\ref{fig:EarthComparison}, right panel). However, these two factors only account for a reduction in the EOS albedo of 0.002-0.004, or, 10\% of the total discrepancy. The remaining part of the discrepancy might be due to differences in the gas opacities. Specifically, if our gas opacity is slightly higher in the visible and in the near infrared, and slightly lower in the mid and far infrared, we would obtain both a higher OLR and a lower TOA albedo\footnote{An increased absorption in the visible/near infrared would reduce the scattered light and thus the TOA albedo. This would also reduce the OLR. However, since the fraction of thermal radiation emitted by the planet in visible/near infrared is small, the OLR reduction in this part of the spectrum can easily be swamped by an OLR increase in mid/far infrared if the absorption in these bands is reduced.}. We are not in the position to test this hypothesis, since it would require accessing the detailed opacity tables used by \cite{yang16}. In any case, we support the EOS results because they can be used to simulate the climate of Earth without further adjustments of the ESTM other variables (see Sec.~\ref{ss:implications}).


\subsubsection{CO$_2$-dominated atmospheric models}
\label{ss:comparisonCO2}

\begin{figure*}[]
\plottwo{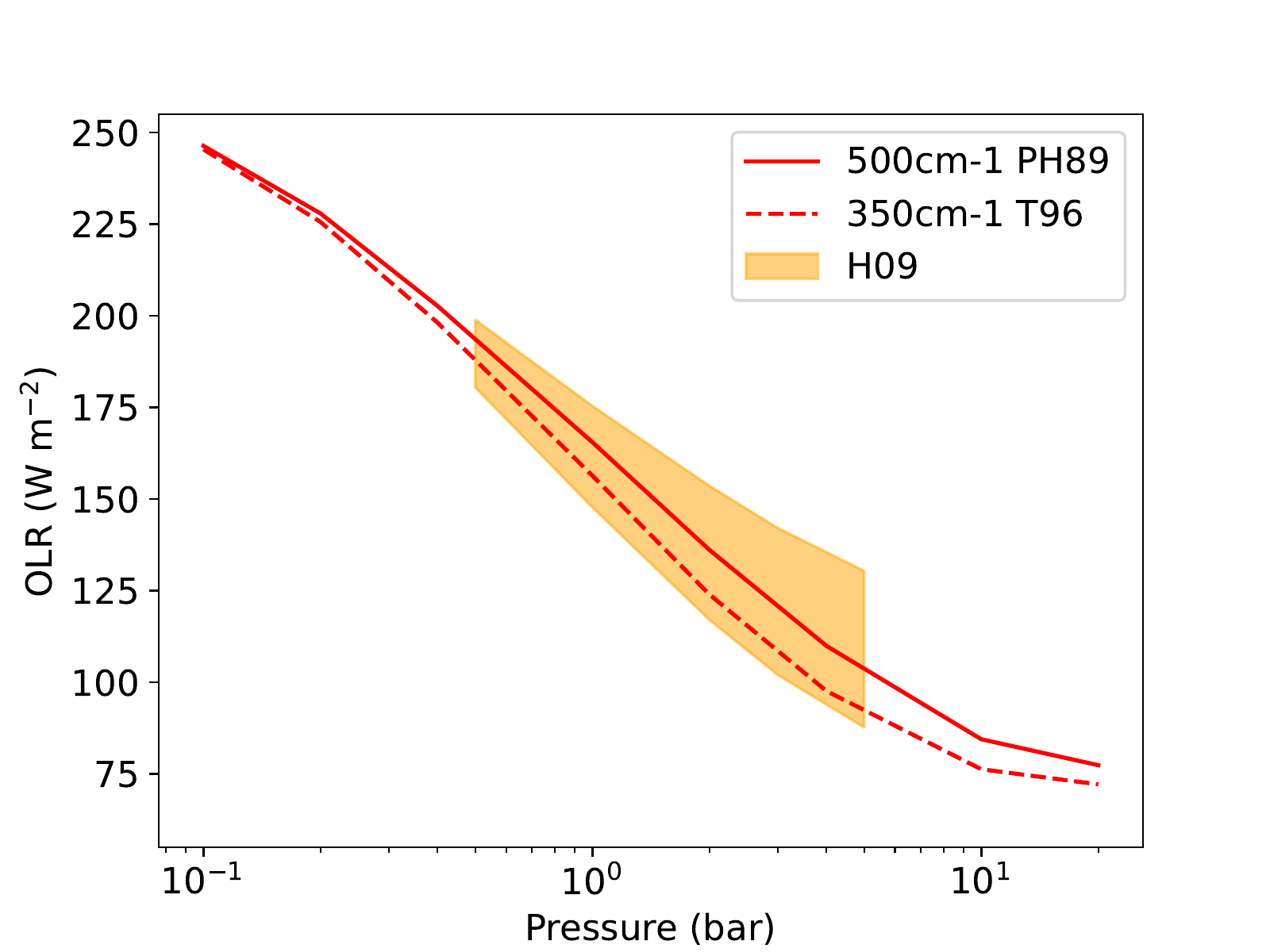}{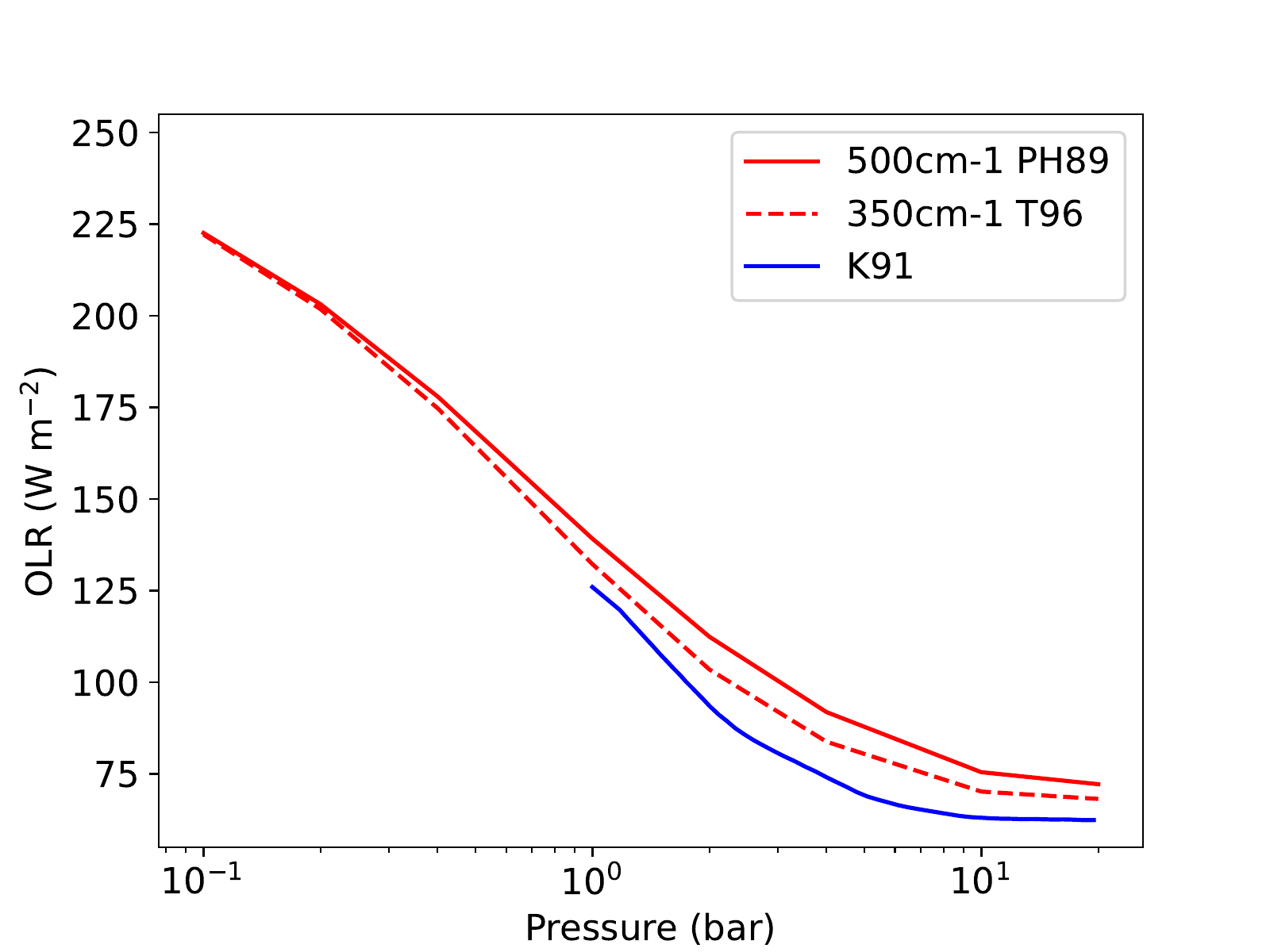}
\caption{The dependence from surface pressure of OLR for dry (left hand panel) and moist saturated (right hand panel) \cdiox -dominated atmospheres. Red lines refer to EOS models. Data for comparison comes from H09 and K91, respectively. \label{fig:CO2Comparison}}
\end{figure*}

In this second round of tests we compared the predictions of the OLR and TOA albedo of \cdiox -dominated atmospheres obtained with EOS with similar predictions 
obtained from other codes. In particular, we tested two cases: the \textit{Dry \cdiox-dominated} one and the \textit{Moist \cdiox-dominated} one.
The vertical profiles are constructed as  explained in Sec.~\ref{ss:co2_lapserate}. 
For consistency with previous published work,
we explored the dependence of the OLR and TOA albedo 
on the \cdiox ~surface partial pressure. In practice, the surface pressure of the dry part (i.e. carbon dioxide) was varied between 0.1 and 20 bar while keeping constant the surface temperature at $T=273$\,K.

In the first test we compared our results with those of H09. These authors  explored: (1)  the effects of different prescriptions  for \cdiox ~in terms of   line profile and  continuum absorption recipes and (2) the forcing  due to addition of various other greenhouse gases (like \water, \methane, N$_2$O etc.). In particular, they calculated the OLR for several pure dry \cdiox ~models with surface pressures in the 0.5-5 bar. In Fig.~\ref{fig:CO2Comparison}, left-hand panel, we compare the results obtained by H09  
with our predictions obtained with PH89 profiles cut at 500 cm$^{-1}$ (solid line) and T96 profiles cut at 350 cm$^{-1}$ (dashed line). Despite some minor differences in the prescriptions adopted, one can see that our predictions 
lie well inside  the envelope of H09 results.    

In the second test we compared our results with those of K91,  
who considered a moist saturated atmosphere dominated by CO$_2$ in the pressure interval 1-34.7 bar. The upper value corresponds to the pressure at which \cdiox ~starts to condense  when $T = 273$\,K at the planetary surface. The comparison against our results is shown in Fig.~\ref{fig:CO2Comparison}, right-hand panel. EOS tends to produce a slightly larger OLR in both of our tested cases  in the entire  pressure interval considered. This is to be expected, since, according to 
\cite{wordsworth10}, K91 model overestimates \cdiox ~opacity due to outdated CIA prescriptions.
The estimated offset is around 14 W m$^{-2}$ for a surface temperature of 250 K, a surface pressure of 2 bars and no water vapor.
%
%
This offset cannot be directly translated to the case under analysis here, because of the presence of water vapor and because of the different surface temperature adopted by \cite{wordsworth10},
but gives a zeroth-order estimate of the magnitude of this effect. The average differences between EOS results and K91 results are equal to 14.7 W m$^{-2}$ for the PH89 line case and to 7.8 W m$^{-2}$ for the T96 case. 

\subsection{Tests with climate models}
\label{ss:implications}

In order to test the reliability of the radiative fluxes obtained with our procedure, the OLR and TOA albedo tables obtained with EOS were applied to calculate the present-day climate of the Earth with the original version \citep{vladilo15} and the upgraded version \citep[][in preparation]{biasiotti21} of the \texttt{ESTM}. At variance with the results presented in the previous sections, the radiative effects of clouds cannot be neglected in this case and as such, they are modeled in both versions of the \texttt{ESTM} by parameterizing the OLR cloud forcing and using a $z$-dependent algorithm for the cloud albedo.
In practice, the cloud OLR forcing is subtracted to the clear-sky OLR calculated by EOS, while the $z$-dependent cloud albedo is included in the surface albedo. Both quantities are evaluated dynamically to account for the cloud coverage fraction. For the specific prescriptions adopted in the \texttt{ESTM}, which are based on experimental data of Earth's clouds, we refer to the relevant papers recalled above.

\begin{figure}[]
\plotone{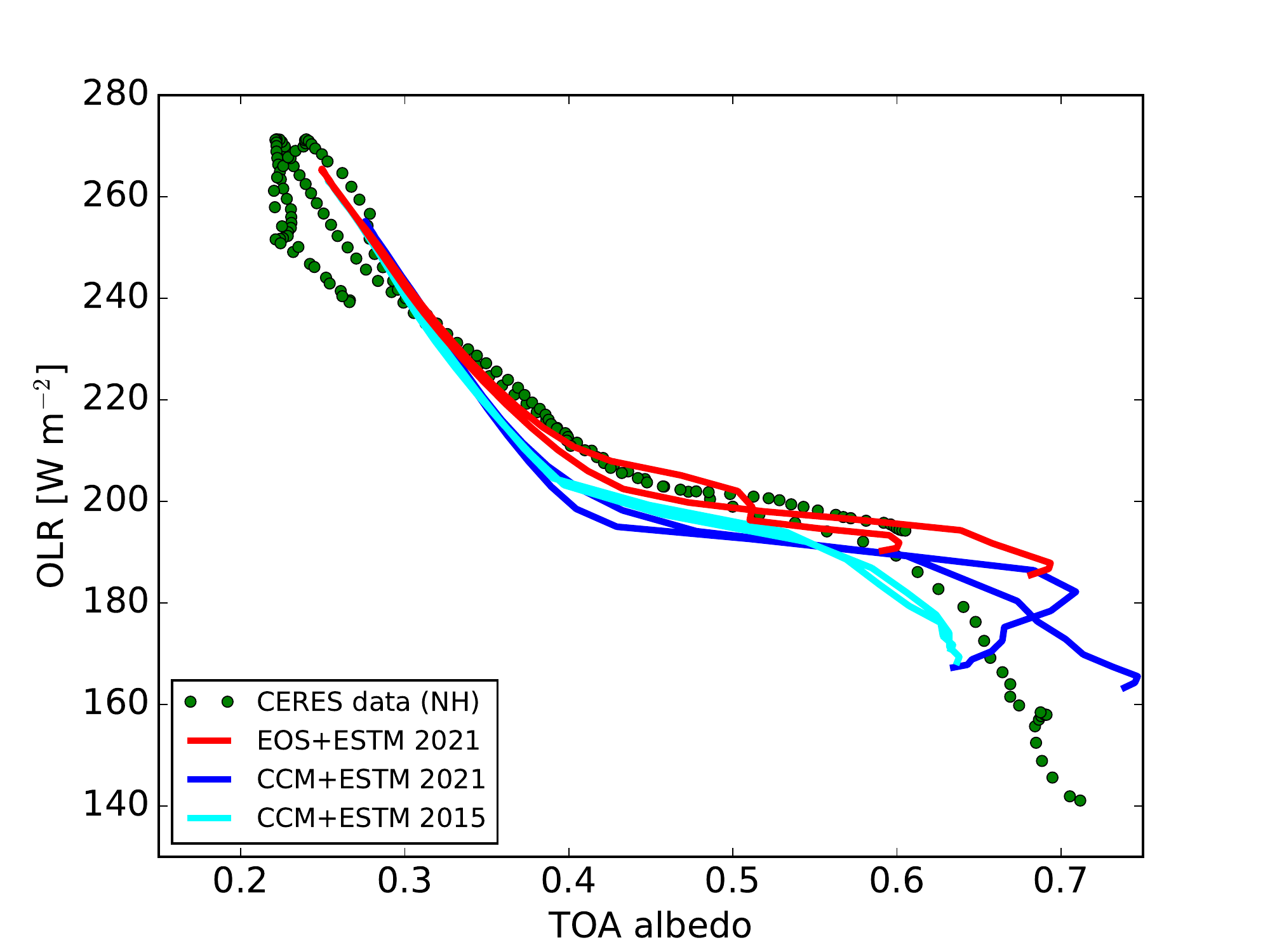}
\caption{
Mean annual values of OLR and TOA albedo of the present-day Earth at different latitudes for  the Northern and  Southern hemispheres. 
Green circles: experimental data obtained with the CERES satellite averaged over the period 2005-2015. 
The solid lines indicate the model predictions for the Earth
obtained adopting: the  EOS (red) or the \texttt{CCM3} (blue) RT  in the new \texttt{ESTM}, and the \texttt{CCM3} RT  in the old \texttt{ESTM} (cyan). The red and cyan lines refer to the  new and old \texttt{ESTM} reference model for the Earth, respectively. See text for details.
\label{fig:OLRvsTOAalb}}
\end{figure}

To test the model predictions we collected the OLR and TOA albedo measurements of the present-day Earth obtained by CERES \citep{loeb18}. For each of these two quantities we extracted the annual mean value as a function of latitude for the period 2005-2015. These experimental data are shown as green circles in Fig.~\ref{fig:OLRvsTOAalb}. Despite the differences between the measurements obtained in the Northern and Southern hemisphere, a clear trend of decreasing OLR with increasing albedo can be seen. This trend is due to the fact that equatorial data have high OLR and low albedo, whereas  polar data have small OLR and high albedo. To perform our test we compared the present-day Earth model calculated using the upgraded \texttt{ESTM} with EOS RT (red lines) and the original \texttt{ESTM} with \texttt{CCM3} RT (cyan lines). One can see that the \texttt{ESTM} with EOS RT provides a better match to the data. Since several climate recipes have been upgraded in the new version of the \texttt{ESTM}, we also considered the results obtained adopting the \texttt{CCM3} RT in the new \texttt{ESTM} (blue lines). Also in this case, the predictions obtained with EOS give a better match to the data.
This test clearly supports the reliability of the RT calculations performed with EOS.

\section{Conclusions} \label{sec:conclusion}

In this work we presented EOS, a procedure for the physical treatment of the vertical RT in the atmospheres of terrestrial-type exoplanets with temperate conditions. EOS is built on the RT code \texttt{HELIOS} and the opacity calculator \texttt{HELIOS-K}, with the addition of physical recipes which are required for treating the chemical composition  and the vertical P-T structure of the atmosphere of habitable worlds. In particular, for the opacities we have considered the continua of absorption of \cdiox~and \water~and the far line wings of \cdiox, and for the atmospheric structure we have considered different definitions of the lapse rate and of the troposphere.
We tested the robustness of EOS against changes of computational variables and atmospheric structure (Sec.~\ref{sec:input}) and we compared its predictions against other radiative transfer codes in the literature (Sec.~\ref{ss:comparison}). Finally, we showed the preliminary results obtained by inserting the TOA fluxes (OLR and albedo) calculated by EOS in a climate energy balance model (\texttt{ESTM}, Sec.~\ref{ss:implications}).
%
%
Our findings can be summarized as follows:
\begin{itemize}
\item EOS is robust to changes of computational parameters (number of layers and the TOA pressure) and responds to changes of input variables in a physically justified manner;
\item the OLR calculated by EOS for the reference \textit{Saturated Modern Earth} atmosphere (Fig.~\ref{fig:EarthComparison}, left-hand panel) is in line with the results obtained by Y16 using a variety of other RT codes. In particular, at low (240-280 K) temperatures the EOS OLR is within the envelope of Y16 results, while at high temperatures (340-360 K) it is slightly (up to $\sim$4\%) higher than the highest Y16 tested model (\texttt{CAM3});
\item the OLR calculated by EOS for the reference \textit{Dry \cdiox-dominated} atmosphere is well within the envelope of the results for different models published by H09 in the tested pressure range (0.5-5 bar, Fig.~\ref{fig:CO2Comparison} left-hand panel). The OLR for the reference \textit{Moist \cdiox-dominated} atmosphere is slightly higher ($\sim$14 Wm$^{-2}$, Fig.~\ref{fig:CO2Comparison} right-hand panel) than that found by K91 in the 1-20 bar range but in line with it when we account for the excess \cdiox ~continuum opacity that the K91 model is claimed to have. We note that both the dry and the moist \cdiox-dominated cases produce compatible results, which support our confidence in the EOS ability to correctly treat these two important greenhouse gases;
\item the TOA albedo calculated by EOS for the \textit{Saturated Modern Earth} atmosphere is slightly lower (typically $\simeq -0.02$) than that predicted by other models (see Fig.~\ref{fig:EarthComparison}, right-hand panel). However, the trend of TOA albedo with respect of temperature, zenith angle, surface albedo and surface pressure are reproduced;
\item once inserted in the \texttt{ESTM} climate model, the TOA fluxes calculated for the \textit{Modern Earth 2} case by EOS correctly reproduce the OLR and TOA albedo observed by the CERES spacecraft. The results obtained using EOS fluxes are in better accordance with observational data than the results obtained using the fluxes calculated by \texttt{CCM3}. Once again, this supports our confidence in the EOS calculations.
\end{itemize}

EOS is integrated, through \texttt{HELIOS-K}, with several online opacity repositories (HITRAN, HITEMP, Kurucz\footnote{\url{http://kurucz.harvard.edu/linelists.html}}...) and is therefore an optimal tool to investigate the RT properties of atmospheres within the wide range of chemical compositions, surface pressures and temperatures that are expected to characterize terrestrial-type exoplanets. Moreover, it is at least one order of magnitude faster than other RT codes, being GPU-enhanced. The fact that EOS is based on codes that are part of the ESP suite adds flexibility in the treatment of several aspects of planetary habitability, including 3D climate modeling \citep[\texttt{THOR},][]{mendonca16,deitrick20} and chemodynamical equilibrium \citep[\texttt{VULCAN} and \texttt{FastChem}, see respectively][]{tsai17,stock18}.

We plan to add continuum opacity and Rayleigh scattering prescriptions for chemical species not yet included in our procedure, such as N$_2$O, and to expand on the treatment of minor gases that could have been important in the early Earth atmosphere. The EOS procedure discloses the possibility to perform calculations of transmission, emission and reflection spectra of the atmosphere of habitable worlds. This will allow us to link the physical state on the planetary surface, evaluated using climate models, to future observations of planetary atmospheric spectra, in a self-consistent way.

\smallskip

We thank the anonymous referee for the careful reading of the manuscript and her/his useful comments.
P.S. wishes to thank the European Space Agency for co-funding his doctoral project (EXPRO RFP IPL-PSS/JD/190.2016).
The authors wish to thank the Italian Space Agency for co-funding the Life in Space project (ASI N. 2019-3-U.0).

\appendix

\section{Description of the procedure}

\begin{figure}
\includegraphics[width=1.00\textwidth]{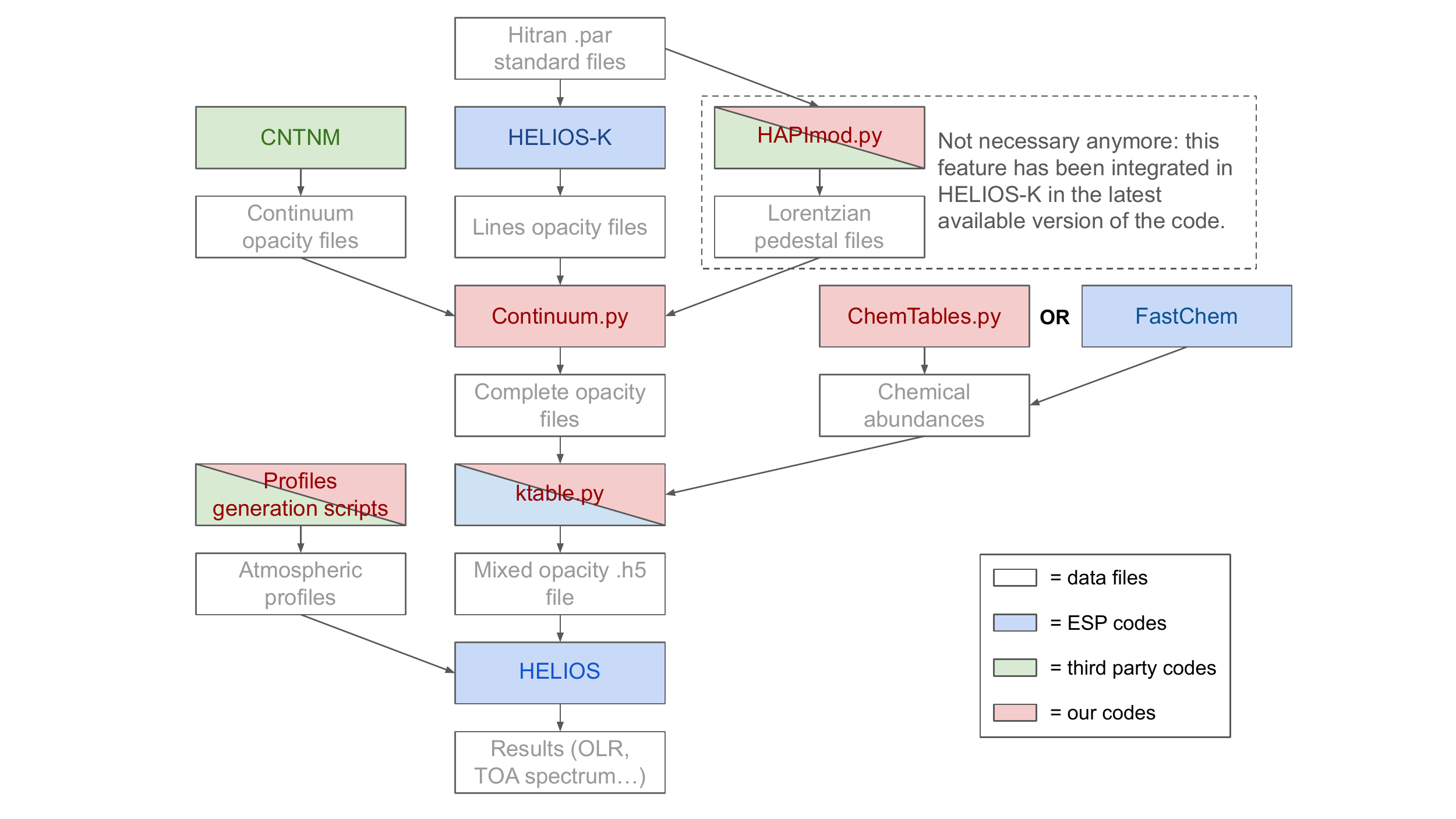}
\caption{The workflow of EOS procedure to generate TOA fluxes for terrestrial-type atmospheres with \texttt{HELIOS}. The color-code is described in the legend of the figure. Double-colored cells refer to codes that have been developed by others but modified by us. The third party codes used are: \texttt{CNTNM} \citep[a module of \texttt{LBLRTM}, see][]{clough05}, \texttt{HAPI} \citep{kochanov16} and the \texttt{MoistAdiabat} class in the CliMT suite by Rodrigo Caballero.}
\label{fig:workflow}
\end{figure}


From a technical standpoint, EOS consists in the steps reported below and shown in Fig.~\ref{fig:workflow}. The physical recipes implemented in our procedure are discussed in the main paper.

As a first step, we download the line parameters and the partition function files from HITRAN \citep{gordon17} for the following species of interest: \water, \cdiox, \methane, \oxy ~and \nitro. These data are then used by \texttt{HELIOS-K} to calculate the absorption coefficients required in the subsequent part of the procedure. Both self- and foreign-broadening due to gas pressure has been evaluated by specifying the correct mixing ratio. Foreign-broadening data in HITRAN are calculated using dry air as background gas. For \water, \methane, \oxy ~and \nitro ~we used Voigt profiles truncated at 25 cm$^{-1}$, while for \cdiox ~we tested three options: (i) a pure Voigt profile truncated at 25 cm$^{-1}$, (ii) a sub-Lorentzian profile using the parameters given by \cite{perrin89} truncated at 500 cm$^{-1}$ and (iii) a sub-Lorentzian profile using the parameters of \cite{tonkov96} truncated at 350 cm$^{-1}$ (as indicated in their paper). When modeling 
Earth-like atmospheres, we used only the first setup, while for \cdiox-dominated atmospheres we tested all three possibilities and adopted the option (ii) as our reference.  

As a second step, we add the \cdiox ~and the \water ~continua to the respective species files, calculated as described in Sec.~\ref{ss:CO2cont} and \ref{ss:H2Ocont}. This is done via the script \texttt{Continuum}. As a technical note, we remark that the \water ~continuum, as calculated by the MT-CKD model, includes the so-called Lorentzian pedestals, which therefore must be subtracted from the absorption coefficient of the lines. The Lorentzian pedestal of a line corresponds to the absorption coefficient of that line at 25 cm$^{-1}$ from its centre. Previous versions of \texttt{HELIOS-K} did not have a built-in option for this subtraction, therefore we modified the \texttt{HAPI} (HITRAN Application Programming Interface)\footnote{\url{https://hitran.org/hapi/}} script to calculate the Lorentzian pedestals from the \texttt{.par} standard line parameters file. We called it \texttt{HAPImod}. The latest version of \texttt{HELIOS-K} integrates this calculation and as such, \texttt{HAPImod} is no longer needed. 
The results obtained via the two different calculations are the same, thus adding robustness to this treatment.

As a third step, we use the \texttt{HELIOS} tool \texttt{ktable} to produce line-by-line mixed opacity tables in the 0-30000 cm$^{-1}$ range and with a $\lambda/\Delta\lambda$ resolution of 3000. We tested two main sets of atmospheres: (i) an Earth-like one, dominated by \nitro ~and \oxy ~and with \cdiox, \methane ~and \water ~as minor components and (ii) a \cdiox-dominated atmosphere with \water ~as a minor component. We then tested several variations of them 
(see Sec.~\ref{sec:input}).
The volume mixing ratios have been calculated on a pressure-temperature grid by the tool \texttt{ChemTables}.


As a fourth step, we launch the main \texttt{HELIOS} code in its \textit{post-processing} mode, in which we specify the vertical pressure-temperature structure of the atmosphere. This is necessary to ensure full control on the input variables (among others, the surface temperature $T_s$) needed to produce the OLR and the TOA albedo tables
in the \textit{reverse procedure}. The vertical structure of the atmosphere is calculated by two sets of codes. For Earth-like atmospheres, we integrate Eq.~\ref{eq:Earthlapserate} using the \texttt{MoistAdiabat} class in the CliMT suite by Rodrigo Caballero, also used in \cite{pierrehumbert11}. 
Instead, for \cdiox-dominated cases, we wrote an ex-novo script that numerically integrated Eq.~\ref{eq:CO2lapserate}.

Finally, as a fifth step, we derived the bulk quantities in terms of TOA fluxes by integrating the results of \texttt{HELIOS} runs, in particular the line-by-line outgoing flux file called \texttt{TOA\_flux\_eclipse}.

\bibliography{biblio}{}
\bibliographystyle{aasjournal}

\end{document}